\documentclass[11pt, a4paper]{article}
\pdfoutput=1
\usepackage{jcappub}

\usepackage{graphicx}%
\usepackage{dcolumn}
\usepackage{bm}
\usepackage{epstopdf}
\usepackage{subfig}
\usepackage{appendix}

\def\Kbar  {\kern 0.2em\overline{\kern -0.2em K}{}\xspace}

\def\Bbar    {\kern 0.18em\overline{\kern -0.18em B}{}\xspace}

\def\Qbar    {\kern 0.08em\overline{\kern -0.08em Q}{}\xspace}

\newcommand{\mev}{\ensuremath{\mathrm{\,Me\kern -0.1em V}}\xspace}
\newcommand{\mevc}{\ensuremath{{\mathrm{\,Me\kern -0.1em V\!/}c}}\xspace}
\newcommand{\mevcc}{\ensuremath{{\mathrm{\,Me\kern -0.1em V\!/}c^2}}\xspace}
\newcommand{\gev}{\ensuremath{\mathrm{\,Ge\kern -0.1em V}}\xspace}
\newcommand{\gevc}{\ensuremath{{\mathrm{\,Ge\kern -0.1em V\!/}c}}\xspace}
\newcommand{\gevcnospace}{\ensuremath{{\mathrm{\,Ge\kern -0.1em V\!/}c}}}
\newcommand{\gevcc}{\ensuremath{{\mathrm{\,Ge\kern -0.1em V\!/}c^2}}\xspace}
\newcommand{\be}{\begin{equation}}
\newcommand{\ee}{\end{equation}}
\newcommand{\benn}{\begin{equation*}}
\newcommand{\eenn}{\end{equation*}}
\newcommand{\bea}{\begin{eqnarray}}
\newcommand{\eea}{\end{eqnarray}}

\def\beq{\begin{equation}}
\def\eeq{\end{equation}}
\def\bea{\begin{eqnarray}}
\def\eea{\end{eqnarray}}

\usepackage{color}

\title{\LARGE Collisions of Jets of Particles from Active Galactic
Nuclei with Neutralino Dark Matter\\[0.25cm]}

\author{\Large Jinrui Huang, Arvind~Rajaraman, and Tim~M.P.~Tait
\\[1.0cm]
\normalsize
{{\it Department of Physics and Astronomy,}}\\
{{\it University of California, Irvine, California 92697}}\\[1.0cm]}
\date{\today\\[1.75cm]}
\emailAdd{jinruih@uci.edu}
\emailAdd{arajaram@uci.edu}
\emailAdd{ttait@uci.edu}

\abstract{
We examine the possibility that energetic Standard Model particles contained in the jets produced by active galactic nuclei (AGN) may scatter off of the dark matter halo which is expected to surround the AGN.  In particular, if there are nearby states in the dark sector which can appear resonantly in the scattering, the cross section can be enhanced and a distinctive edge feature in the energy spectrum may appear.  We examine bounds on supersymmetric models which may be obtained from the Fermi Gamma-ray Space Telescope observation of the nearby AGN Centaurus A.}


\begin{document}
\maketitle
\section{Introduction}
\label{sec:intro}

Active galactic nuclei (AGN)~\cite{Schlickeiser:1994aa, Mastichiadis:1994ab, Bottcher:1996hn, Dar:1996qv, Dermer:1993cz, Mannheim:1989kp, Gaisser:1994yf} are among the most powerful natural accelerators in the Universe, producing jets of charged particles which travel macroscopic distances and can even be imaged from the Earth through the subsequent production of high energy photons. While the nature of the jet constituents (proton or electron) remains a subject of some controversy, there is no doubt that they are among the highest energy and most interesting objects in the cosmos, with energies reaching into tens of GeV and with huge luminosities.

The supermassive black holes which power these jets are typically hosted in very large galaxies, where one would
expect the largest concentrations of dark matter (DM) also occur.  Taken together, the combination of dense dark matter with high energy particles represent a new opportunity to study the interactions of dark matter and Standard Model (SM) particles at high energies. Indeed, while particle accelerators on the Earth reach higher energies and with (much) better understood beams, the huge density of dark matter expected close to AGNs implies that they represent a unique opportunity to learn about dark matter interactions with high energy SM particles.

If there is some other particle in the ``dark sector" which can be produced as a resonance in the scattering of dark matter with a jet particle, the rate for scattering in the high energy tail of the jet may be enhanced.
Beam particles with energies $E$ greater than,
\bea
E & \geq & \frac{1}{2} \frac{m_{\tilde{e}}^2 - m_{\tilde{\chi}}^2}{m_{\tilde{\chi}}}
\eea
(where $m_{\tilde{\chi}}$ is the mass of the dark matter and $m_{\tilde{e}}$ the mass of the resonance)
have sufficient energy to produce the resonance on-shell, leading to an enhancement of the cross section. Since
the beam particles are presumably charged (so that they can be efficiently accelerated by the AGN), the resonant state should also be charged.  As a result, its decays can produce gamma rays which can stick out from the background produced by the AGN and jet themselves. Such a situation is common in the most popular theories of dark matter, such as in supersymmetric theories in which a neutralino plays the role of dark matter \cite{Goldberg:1983nd}, and which can be up-scattered by an energetic beam electron into a selectron \footnote{Similarly, in theories with Universal Extra Dimensions~\cite{Appelquist:2000nn}, the dark matter particle may be a Kaluza-Klein excitation of the hypercharge vector boson \cite{Servant:2002aq,Cheng:2002ej}, which may up-scatter into a Kaluza-Klein mode of the electron \cite{Gorchtein:2010xa}.} and was first proposed by  \cite{Bloom:1997vm} and recently studied in~\cite{Gorchtein:2010xa}. Such a degeneracy may even be central to realizing a neutralino WIMP as dark matter, to allow for efficient enough coannihilation to produce the observed density from thermal freeze-out. In fact, a very degenerate selectron is challenging to detect at the LHC, because its visible decay products have energies controlled by $m_{\tilde{e}} - m_{\tilde{\chi}}$, and are typically very soft. Thus, the signal from AGNs is enhanced precisely where the LHC has difficulty, and can help study motivated but otherwise experimentally difficult regions of parameter space~\cite{Andreev:2004qq}.

Since the resonant state must be charged, its subsequent de-excitation includes higher order processes in which a photon is produced in addition to the primary SM and dark matter decay products. While suppressed by $\alpha$, such processes may be enhanced by collinear logarithms that rise when the photon is emitted along the same
direction as the final state charged SM particle. Unlike charged particles, these radiated photons have very long mean free paths, and could be detected by detectors on the Earth as an indicator that the jet-DM scatterings are taking place. For dark matter masses on the order of a few hundred GeV, these photons fall precisely in the range for which the Fermi Gamma-ray Space Telescope~\cite{Atwood:2009ez} is sensitive, and thus it is precisely the right time to explore such signals. In fact, as pointed out in Ref.~\cite{Gorchtein:2010xa}, the rise in cross section as beam particles cross the $\tilde{e}$ threshold leads to a very interesting feature in the spectrum of gamma rays, a rise in the number of high energy photons followed by a sharp edge.  This feature provides an interesting discriminant from the background of gamma rays emitted from the jet itself, one which conventional astrophysical sources are expected to have difficulty faking. Inspired by Ref.~\cite{Gorchtein:2010xa}, we study the dark matter and AGN electron jet scattering process in SUSY models, presenting the complete scattering amplitudes and performing an explicit check of gauge invariance. We scan the parameter space of neutralino and selectron masses to define precise constraints from the Fermi-LAT data.

This article is organized as follows. In Section~\ref{sec:astro}, we discuss the astrophysical inputs to the
computation, namely the density profile of dark matter around the AGN and the intensity and energy spectrum of the beam particles. In Section~\ref{sec:xsec}, we revisit the computation of the cross sections for scattering in supersymmetric theories.  In Section~\ref{sec:susy} we examine the bounds one can place for different astrophysical assumptions. We conclude in Section~\ref{sec:outlook}.

\section{Astrophysical Inputs}
\label{sec:astro}

In this section, we discuss the astrophysical inputs to our estimates. These boil down into the properties of particles inside the AGN jet, and the density of dark matter surrounding it.  In both cases, we follow closely the discussion of Ref.~\cite{Gorchtein:2010xa}, which contains a more detailed discussion, and study of the impact varying these parameters has on the signal.  We restrict ourselves to looking at Centaurus A, which offers a good balance of being relatively close by ($D \sim 3.7$~Mpc \cite{Ferrarese:2006ke}) as well as a massive object surrounded by a dense halo of dark matter.

\subsection{Dark Matter Distribution}
\label{subsec:DMdist}
The impact of the dark matter distribution is through its integral along the
``line of sight" of the jet particles,
\bea
\delta_{\rm DM} & \equiv  & \int^{r_{\rm max}}_{r_{\rm min}} dr~\rho (r)
\eea
where $\rho (r)$ is the density of dark matter at a radial distance $r$ from the center of the AGN, $r_{\rm min}$ characterizes the minimum distance of interest (the base of the jet), and $r_{\rm max}$ the distance at which the jet loses cohesion and becomes irrelevant.  In practice, results depend sensitively on $r_{\rm min}$ and are rather insensitive to $r_{\rm max}$ \cite{Gorchtein:2010xa} due to the fact that dark matter density distribution $\rho(r)$ drops steeply as the the radius r increases.

The dark matter density distribution surrounding
a compact object was modeled in Ref.~\cite{Gondolo:1999ef} in the collisionless
limit and under the assumption that the central black hole grew adiabatically
by accretion.  The resulting steady-state distribution
is characterized by a central spike of dark matter, which gets depleted in the
core by pair annihilation.  The resulting distribution is thus sensitive to the
underlying particle physics model of dark matter, with the central density
enhanced for WIMPs with very low (or in the case of a purely asymmetric
dark matter halo, zero) annihilation cross sections.  Ref~\cite{Gondolo:1999ef}
begins with a cusped density profile
$\rho_{\rm i} (r) = \rho_0 (r / r_0)^{-\gamma}$,
and determines a final profile (close to the center) of the
form,
\bea
\rho (r) & = & \frac{\rho^\prime (r) ~\rho_{\rm core}}
{\rho^\prime (r) + \rho_{\rm core}}~,
\eea
where the core density is given in terms of the velocity-averaged annihilation rate
$\langle \sigma v \rangle$ and the age of the black hole $t_{\rm BH}$
as,
\bea
\rho_{\rm core} & \simeq & \frac{m_{\tilde{\chi}}}
{ \langle \sigma v \rangle ~ t_{\rm BH} }~.
\eea
The dependence on $r$ is contained in the function,
\bea
\rho^\prime (r) & = & \rho_0 \left( \frac{R_{\rm sp}}{r_0} \right)^{-\gamma}
\left( 1 - \frac{4 R_S}{r} \right)^{3}
\left( \frac{R_{\rm sp}}{r} \right)^{\gamma_{\rm sp}}~,
\eea
where $\gamma_{\rm sp}$ is slope of the density profile in the dark matter
spike, $R_{\rm sp}$ its radius, and $R_S$ is the Schwarzschild radius.  Below
$4 R_S$, the dark matter is accreted into the black hole, and
has $\rho = 0$.  The spike radius and power are given by,
\bea
R_{\rm sp} = \alpha_\gamma ~ r_0 ~
\left( \frac{M_{\rm BH}}{\rho_0 ~ r_0^3} \right)^{\frac{1}{3 - \gamma}}
& ~~{\rm and}~~ &
\gamma_{\rm sp} = \frac{9 - 2 \gamma}{4 - \gamma}~,
\eea
where $\alpha_\gamma \sim 0.293 \gamma^{4/9}$
for $\gamma \ll 1$ ~\cite{Gondolo:1999ef}.
This analysis neglects the influence of stellar populations close to the
the supermassive black hole, which can gravitationally scatter dark matter
out of the spike, resulting in a somewhat shallower profile \cite{Gnedin:2003rj}.

Following Ref.~\cite{Gorchtein:2010xa}, we normalize the distribution
such that it falls within the uncertainty on the measured mass of the black hole itself.
The mass of the black hole is determined by measurements of stellar
kinematics to be  \cite{Neumayer:2010kw},
\bea
M_{\rm BH} = (5.5 \pm 3.0) \times 10^7 M_{\odot}
\eea
which corresponds to $R_S  \simeq 5 \times 10^{-6} ~{\rm pc}$.  Given
a choice of $\gamma$, we determine $\rho_0$ by requiring,
\bea
3 \times 10^7 M_\odot & \geq &
\int_{4 R_S}^{10^5 R_S} dr ~ 4\pi r^2 ~ \rho (r)~,
\eea
where $10^5 R_S$ characterizes the distances at which the stellar populations
are measured.

Ultimately, the factor $\delta_{\rm DM}$ is very sensitive to the product
$\langle \sigma v \rangle \times t_{\rm BH}$, which determines the density within
the spike.  For black hole ages between $10^8$ to $10^{10}$ years and cross
sections ranging from $10^{-26}$~cm$^3$ s$^{-1}$ (appropriate
for a thermal relic) to $10^{-30}$~cm$^3$ s$^{-1}$ (typical for the
``coannihilation" region of the MSSM, in which the superpartners of the charged
leptons play an active role in freeze-out), one finds
$\delta_{\rm DM}$ ranging from,
\bea
\delta_{\rm DM} \sim 10^8 ~~{\rm to}~~ 10^{11}~~ {M_\odot}~ / ~{{\rm pc}^3}
\eea
with rather mild
dependence on the choices
of $\gamma$, $r_{\rm min}$, and $r_0 \sim r_{\rm max} \sim 15$~kpc
\cite{Gorchtein:2010xa}.

\subsection{AGN Jet Particles}
\label{subsec:AGNJet}

For our purposes, the details of the AGN jet geometry
are not very important, while modeling the energy distribution of particles in
the jet is crucial.  We assume most of the particles in the Centaurus A jet
are electrons.  Based on observation of emitted gamma rays by the
Fermi LAT experiment~\cite{:2010fk}, we assume the distribution in
the electron boost is a broken power law,
\begin{equation}
\label{eqn:JetEne}
\frac{d \Phi_{e}^{\rm (AGN)}}{d \gamma^{\prime}} (\gamma^{\prime})
~ = ~\frac{1}{2} k_e \gamma^{\prime \, -s_1}
\left[1 + (\frac{\gamma^{\prime}}{\gamma^{\prime}_{br}})^{(s_2 - s_1)} \right]^{-1}
~~~~~~~~~~~~~
\left(\gamma^{\prime}_{\rm min}
< \gamma^{\prime} < \gamma^{\prime}_{\rm max}\right) \;.
\end{equation}
Here the primed variables refer to the ``blob frame", in which the
electrons move isotropically.
The parameters $s_1$, $s_2$, $\gamma^{\prime}_{br}$,
$\gamma^{\prime}_{\rm min}$
and $\gamma^{\prime}_{\rm max}$ are fixed to~\cite{:2010fk}
\bea
s_1 = 1.8,
\qquad s_2 = 3.5,
\qquad \gamma_{br}^{\prime} = 4 \times 10^5,
\qquad \gamma_{\rm min}^{\prime} = 8 \times 10^2,
\qquad \gamma_{\rm max}^{\prime} = 10^8 \; .
\eea
The normalization $k_e$ can be determined from the jet power in
electrons, which is defined in the black hole frame as,
\bea
\label{eqn:JetPower}
L_e & = & \int_{-1}^{1} ~\frac{d \mu}{\Gamma_B (1 - \beta_B \mu)} ~
\int_{\gamma_{\rm min}}^{\gamma_{\rm max}} ~
d \gamma  (m_e \gamma) ~~
\frac{d \Phi_{e}^{AGN}}{d \gamma}
\biggl(\gamma(\Gamma_B (1 - \beta_B \mu))\biggr) \;, \\
\eea
where $\mu = \cos \theta$ ($\theta$ is the polar angle with respect to the jet axis)
and $\gamma = \epsilon/m_{e}$ where
$\epsilon$ is the energy of the electron.
Similarly, the primed variables in the blob frame are
$\mu^{\prime} = \cos \theta^{\prime}$ and
$\gamma^{\prime} = \epsilon^{\prime}/m_{e}$. The quantities in the two frames
are related by the blob velocity $\beta_B$ and boost
$\Gamma_B = 1/\sqrt{1 - \beta_B^2}$ ( $\sim 3$ for Centaurus A) as,
\bea
\mu^{\prime} = \frac{\mu - \beta_B}{1 - \beta_B~ \mu}
\; , \qquad
\gamma^{\prime} = \gamma ~\Gamma_B ~(1 - \beta_B \mu) \;,
\eea
or equivalently,
\bea
\mu = \frac{\mu^{\prime} + \beta_B}{1 + \beta_B \mu^{\prime}}
\; , \qquad
\gamma = \frac{ \gamma^{\prime}}{\Gamma_B ~(1 - \beta_B \mu)} \;.
\eea
The limits on the $\gamma$ integral are given by,
\bea
\gamma_{\rm min} = \frac{\gamma^{\prime}_{\rm min}}
{\Gamma_B~(1 - \beta_B \mu)}
\; , \qquad
\gamma_{\rm max} =  \frac{\gamma^{\prime}_{\rm max}}
{\Gamma_B~(1 - \beta_B \mu)} \; ,
\eea
whereas the limits on $\mu$ in the black hole frame correspond to
$0.9 - 1$, leading to a highly collimated jet of electrons.

The function
${d \Phi_{e}^{\rm (AGN)}}/{d \gamma}$
can be written explicitly as,
\begin{equation}
 \frac{d \Phi_{e}}{d \gamma}^{\rm (AGN)}
 \biggl( \gamma(\Gamma_B (1 - \beta_B \mu))\biggr)
 = \frac{1}{2} k_e \biggl[\gamma \Gamma_B (1 - \beta_B \mu)\biggr]^{-s_1}
 \biggl[1 + \biggl(\gamma (\Gamma_B (1 - \beta_B \mu))/\gamma_{br}^{\prime}
  \biggr)^{s_2 - s_1} \biggr]^{-1} \;.
\end{equation}
Saturating the upper limit of $L_e \leq 10^{46}$ erg $s^{-1}$
\cite{Gorchtein:2010xa}, we arrive at
\bea
\frac{1}{2}k_e = 4.2088 \times 10^{49} s^{-1}~.
\eea

\section{Cross Sections}
\label{sec:xsec}

\begin{figure}
\centering
\subfloat{\includegraphics[scale=0.7]{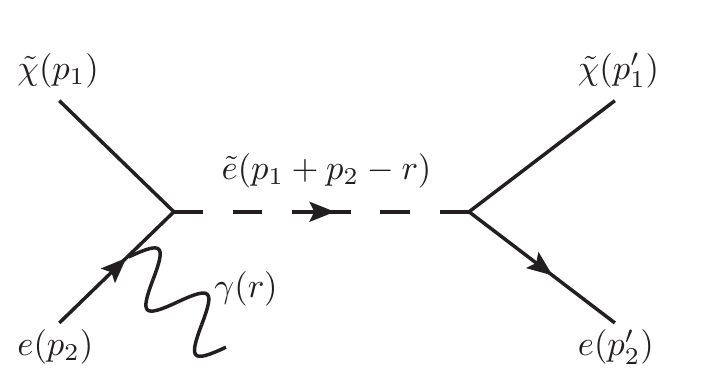} \label{fig:FeynDiagSch-a}}
\subfloat{\includegraphics[scale=0.7]{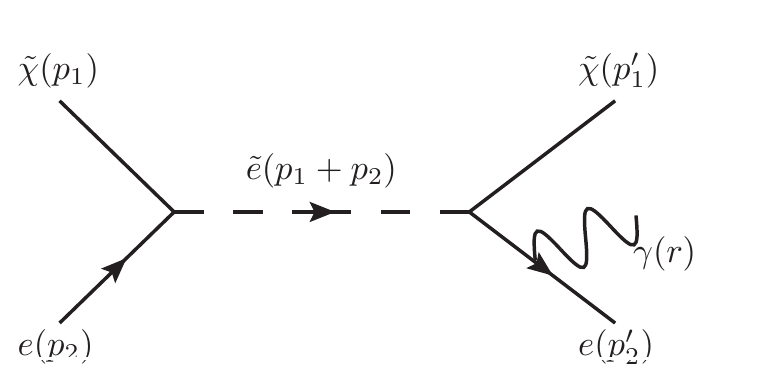} \label{fig:FeynDiagSch-b}}
\subfloat{\includegraphics[scale=0.7]{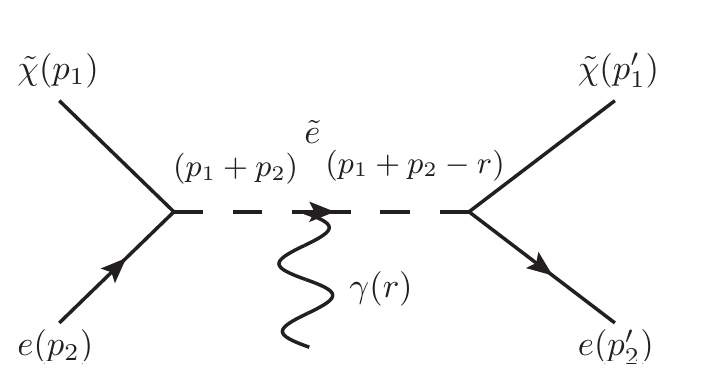} \label{fig:FeynDiagSch-c}}\\
\subfloat{\includegraphics[scale=0.7]{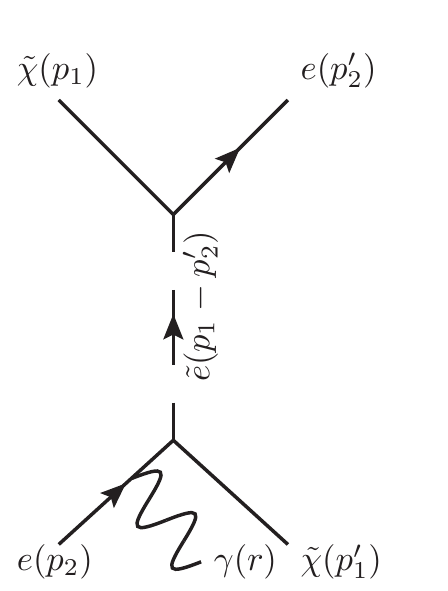} \label{fig:FeynDiagUCh-a}}
~~~~~~~~~~~~
\subfloat{\includegraphics[scale=0.7]{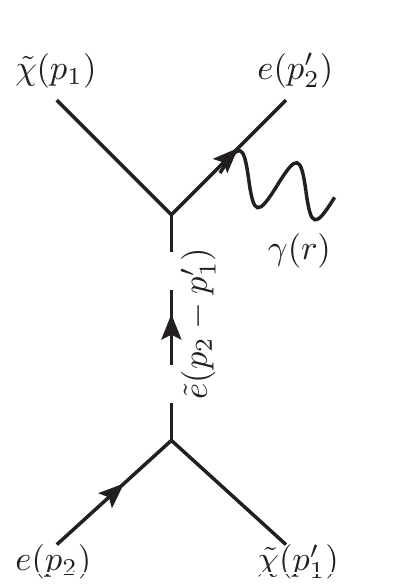} \label{fig:FeynDiagUCh-b}}
~~~~~~~~~~~~
\subfloat{\includegraphics[scale=0.7]{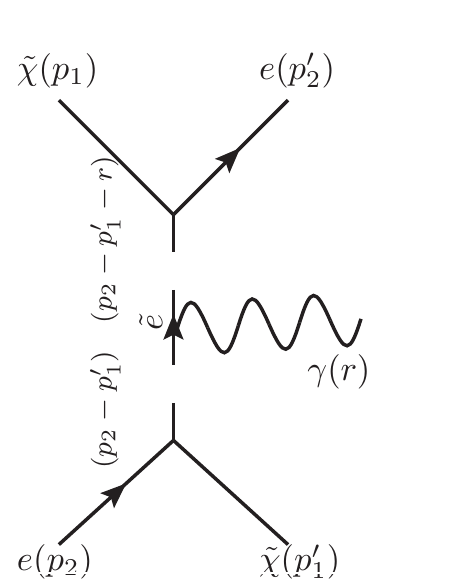} \label{fig:FeynDiagUCh-c}}
\caption{Feynman diagrams for the process
$e^{-} ~ \tilde{\chi} \rightarrow e^{-} ~ \tilde{\chi} ~ \gamma$ through
$\tilde{e}$ exchange.  The amplitudes corresponding to these diagrams are,
sequentially, $M_1 - M_6$.}
\label{fig:FeynDiagSch}
\end{figure}

The total cross section for the process
$e ~ \tilde{\chi} \rightarrow e ~ \tilde{\chi} ~ \gamma$ may be written,
\begin{equation}
d \sigma = (2\pi)^4 \delta^4(p_1 + p_2 - p_1^{\prime} - p_2^{\prime} - r)
\frac{1}{4 m_{\tilde{\chi}} E_2}
\frac{d^3 p_{1}^{\prime}}{(2 \pi)^3 2E_1^{\prime}}
\frac{d^3 p_2^{\prime}}{(2 \pi)^2 2E_2^{\prime}}
\frac{d^3 r}{(2 \pi)^3 2E_{\gamma}} \sum |\mathcal{M}|^2 \;,
\end{equation}
where the $\sum |\mathcal{M}|^2$ is the sum of all
of the matrix elements squared.
In addition to the $s$-channel contributions included
in~\cite{Gorchtein:2010xa}), the $u$-channel $\tilde{e}$ exchange diagrams
are also included, as shown in the Fig.~\ref{fig:FeynDiagSch}.
We ignore $t$-channel contributions mediated by the SM $Z$ boson as it is
non-resonant, and thus negligible. The amplitudes corresponding to these
diagrams are labelled $M_1 - M_6$.

The experimental observable is the distribution of photon energies,
$\nu S_{\nu} \equiv E_{\gamma}^2 \frac{d\Phi_{\gamma}}{dE_{\gamma}}$ at
a fixed angle relative to the jet axis ($\sim 68^{\circ}$ for the Centaurus A).
For fixed incoming electron energy, we integrate over the phase space of the
final state neutralino and electron, resulting in
${d^2 \sigma}/{d E_{\gamma} d \Omega_{\gamma}}$.  We work in the rest frame of the initial neutralino dark matter particle, which is expected to be moving
non-relativistically ($\sim 300$~km/s \cite{Karachentsev:2004hg})
in the rest frame of the Earth.
${d^2 \sigma}/{d E_{\gamma} d \Omega_{\gamma}}$ can be expressed,
\begin{equation}
\frac{d^2 \sigma}{d E_{\gamma} d \Omega_{\gamma}}
= \frac{1}{(2 \pi)^2} \frac{1}{4 m_{\tilde{\chi}} E_2} \frac{1}{8E_1^{\prime}}
\int d \Omega_{p_2^{\prime}}
\biggl(E_2^{\prime} E_{\gamma} \frac{1}{|1 + J|} \sum |\mathcal{M}|^2 \biggr),
\end{equation}
where $J$ is a Jacobian factor from the integration of the function
$\delta^4(p_1 + p_2 - p_1^{\prime} - p_2^{\prime} - r)$ over $d E_2^{\prime}$ and can be explicitly written as
\bea
J & = & \frac{E_2^{\prime}}{E_1^{\prime}} - \frac{(\vec{p}_1 + \vec{p}_2 - \vec{r}) \cdot \vec{p}_2^{\prime}}{E_1^{\prime} E_2^{\prime}} \\
& = & \frac{E_2^{\prime}}{E_1^{\prime}} - \frac{(p_{2x} - r_{x})p_{2x}^{\prime} + (p_{2y} - r_{y})p_{2y}^{\prime} + (p_{2z} - r_{z})p_{2z}^{\prime}}{E_1^{\prime}E_2^{\prime}}.
\eea

The full expression for the amplitude squared may be found in
Appendix~\ref{app:amp}.
The cross section is enhanced in three kinematic configurations:
\begin{itemize}
\item $M_{1-3}$ are resonantly enhanced when the incoming electron energy
is such that the intermediate selectron is approximately on-shell
\cite{Bloom:1997vm}.
\item $M_2$ and $M_5$ have
a collinear enhancement when the emitted photon lines
up with the final state electron \cite{Gorchtein:2010xa}.
\item $M_2$ and $M_5$ have a soft enhancement for low energy photon emission.
\end{itemize}
Note that because the jet axis makes a fixed angle with respect to the line of
sight from the Earth, there is no possibility of a collinear enhancement from
the initial electron in $M_1$ and $M_4$.  In practice, the soft enhancement
(the infinity of which is formally matched by the one-loop QED correction to
$e^- ~ \tilde{\chi} \rightarrow e^- ~ \tilde{\chi} $ scattering) is not
observationally interesting.  The dominant contribution comes from the on-shell
selectron, collinear photon region of $M_2$:
\bea
\label{eqn:AmpSquM2}
|M_2|^2 & = &
\frac{\alpha_{\mbox{\tiny{eff}}}~~(p_1^{\prime} \cdot r)(p_1 \cdot p_2)}
{|\Sigma_{s}|^2 ~ \biggl((p_2^{\prime} \cdot r) - \frac{m_{e}^2}{2} \biggr)}
 \;,
\eea
where
$\alpha_{\mbox{\tiny{eff}}} = 4 \pi \alpha_{\mbox{\tiny{EM}}}
(a_{\mbox{\tiny{L}}}^4 + a_{\mbox{\tiny{R}}}^4)$ is the effective
neutralino-selectron-electron
coupling.  We assume the left- and right-handed selectrons are degenerate in
mass, and a mostly bino neutralino, for which,
\bea
a_{\mbox{\tiny{R}}}^2 & = & 2g_w^2 \tan\theta_w^2~, \\
a_{\mbox{\tiny{L}}}^2 & = & a_{\mbox{\tiny{R}}}^2/4~.
\eea
The factor
$1/|\Sigma_{s}|^2$ is the selectron propagator, which has been modified
to maintain  gauge invariance\footnote{Proof of
gauge invariance (correcting Ref~\cite{Gorchtein:2010xa}) can be found in
Appendix~\ref{app:gauInv}.},
using the prescription of Ref.~\cite{Baur:1995aa}.
To avoid the divergence from soft radiation, $(p_2^{\prime} \cdot r)$ in the
propagator has been shifted by ${m_{e}^2}/{2}$.

\begin{figure}[tbh]
\includegraphics[width=8.2cm,angle=0]{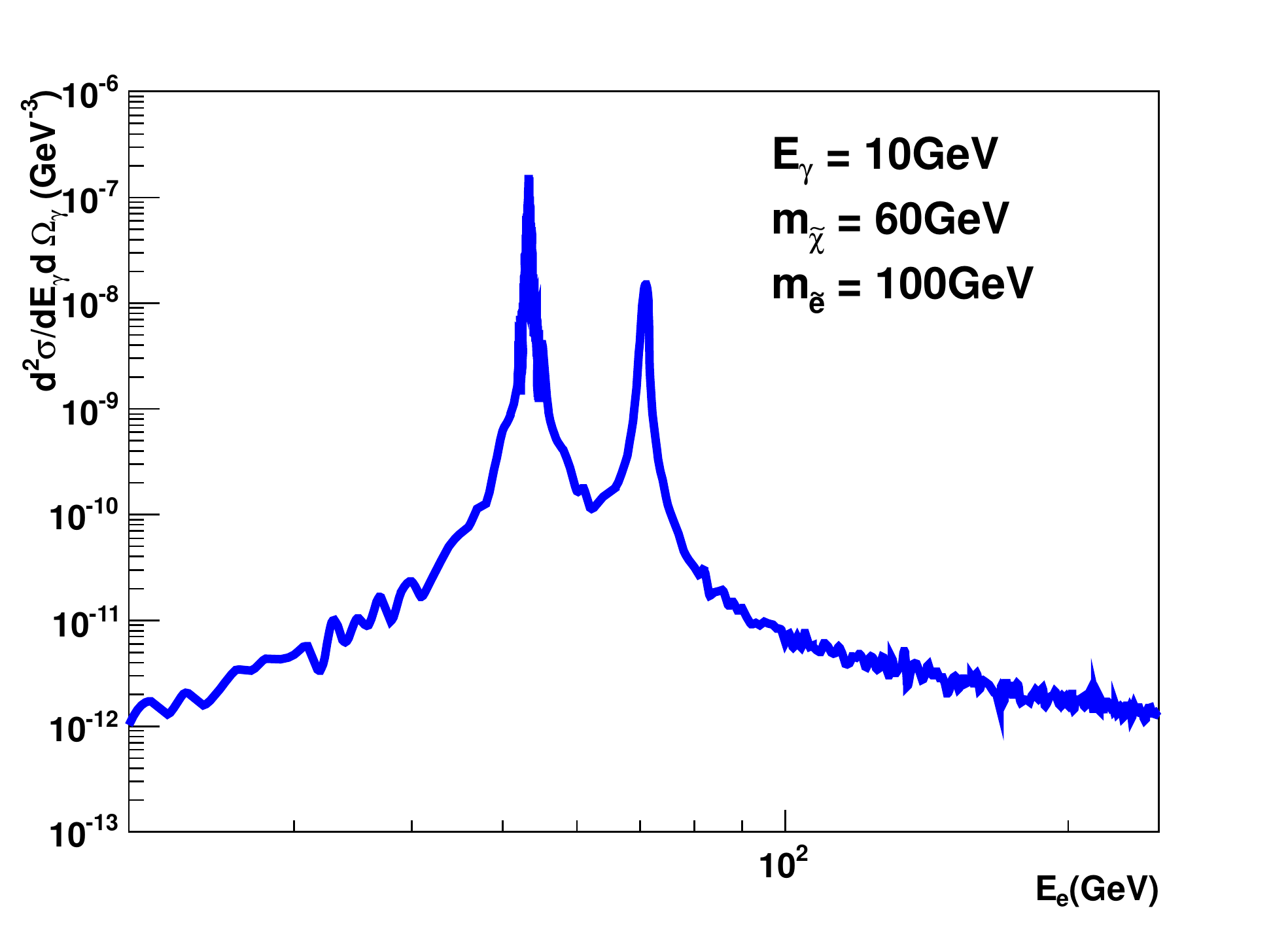}
\includegraphics[width=8.2cm,angle=0]{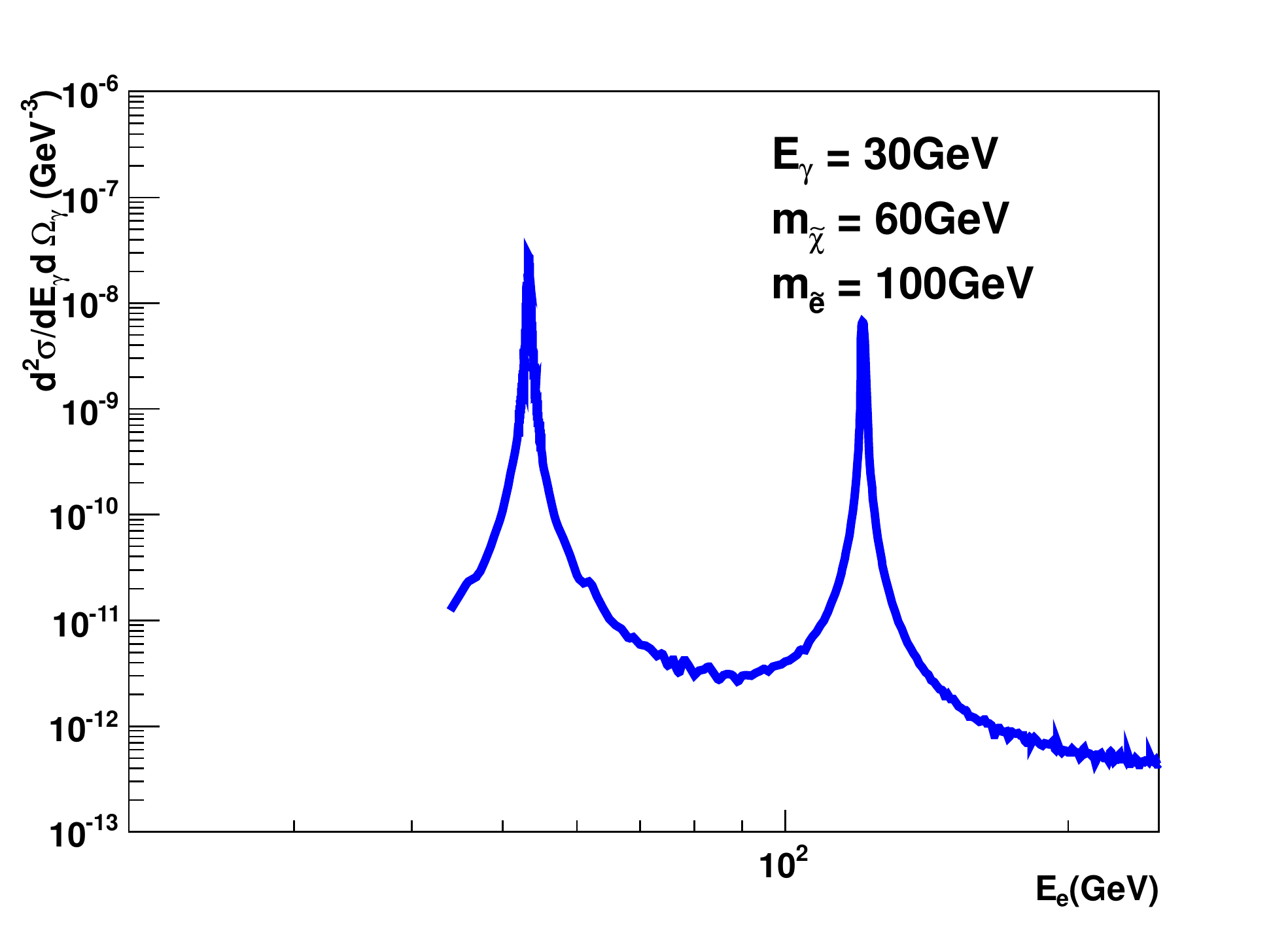}
\includegraphics[width=8.2cm,angle=0]{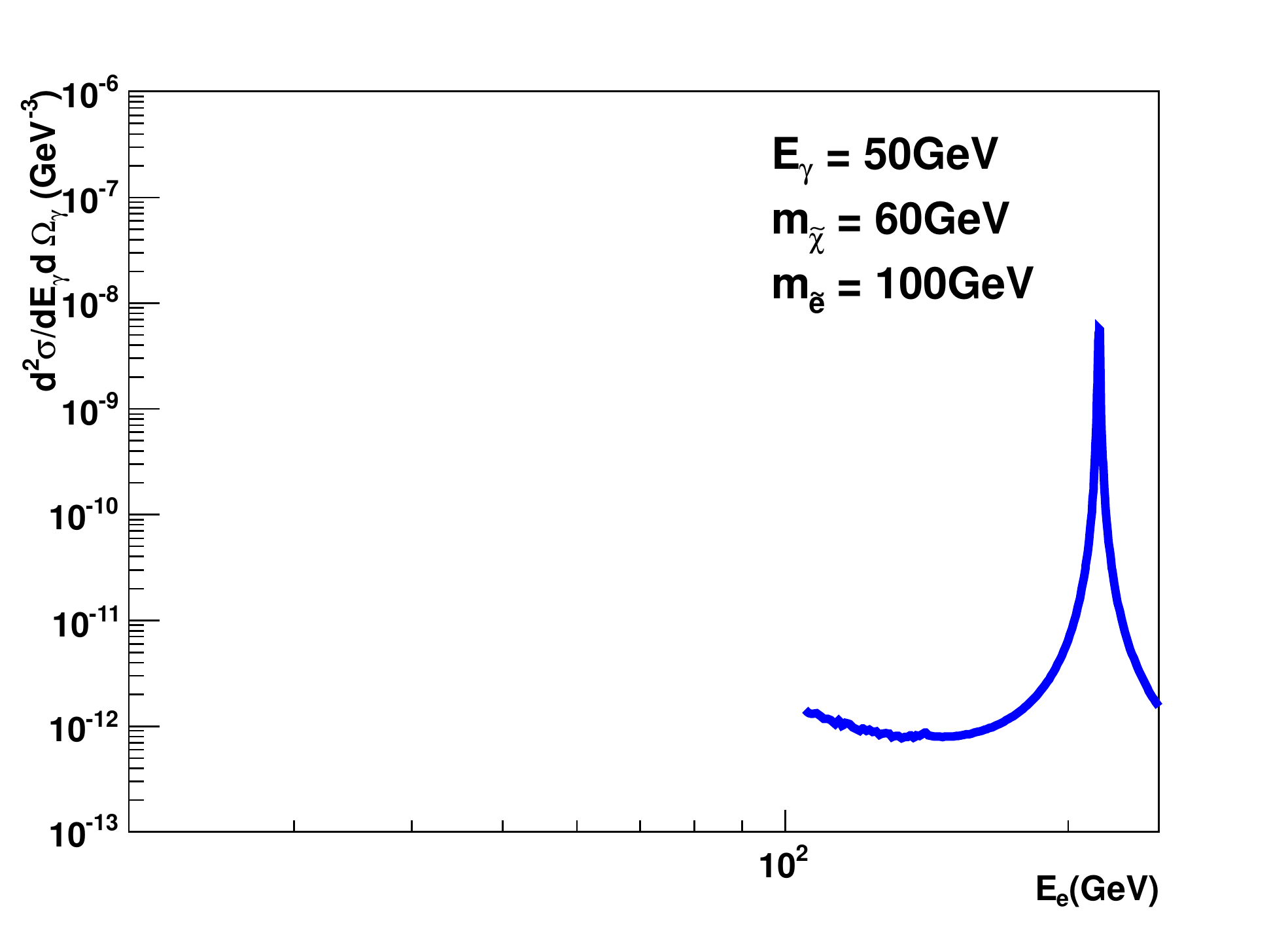}
\caption{\label{fig:dsigdE}
Differential cross sections, $d^2 \sigma / dE_{\gamma} d\Omega_{\gamma}$ for the process
$e^- \tilde{\chi} \rightarrow e^- \tilde{\chi} \gamma$, for a neutralino
mass of 60 GeV and selectron mass of 100 GeV, and fixed
photon energies (top to bottom) of 10~GeV, 30~GeV, and 50~GeV.
}
\end{figure}

The differential cross sections $d^2\sigma / dE_e d\Omega_\gamma$
are shown as a function of the incoming
electron energy in Fig.~\ref{fig:dsigdE}, for fixed photon energies of 10 GeV,
30 GeV and 50 GeV. As a benchmark point, we have chosen the mass of the
neutralino to be $m_{\tilde{\chi}} = 60$ GeV and the mass of the
selectron to be $m_{\tilde{e}} = 100$ GeV.
Notice that for a photon energy of 10 GeV or 30 GeV,
there are two peaks which arise from the (shifted) resonant behavior of
$M_2$ and $M_1$ separately.
However, as the photon energy increases, the incoming electron energy must
be large enough to produce such a hard photon in the scattering process,
closing off the possibility for scattering (for fixed $E_\gamma$) below
a certain threshold.  Once the threshold rises above $\sim m_{\tilde{e}} / 2$,
it becomes impossible to produce a final state photon from an on-shell
$\tilde{e}$ decay, and the $M_2$ resonance effectively disappears.

\section{Bounds on Supersymmetric Parameter Space}
\label{sec:susy}

\begin{figure}
\includegraphics[width=10.80cm, angle=0]{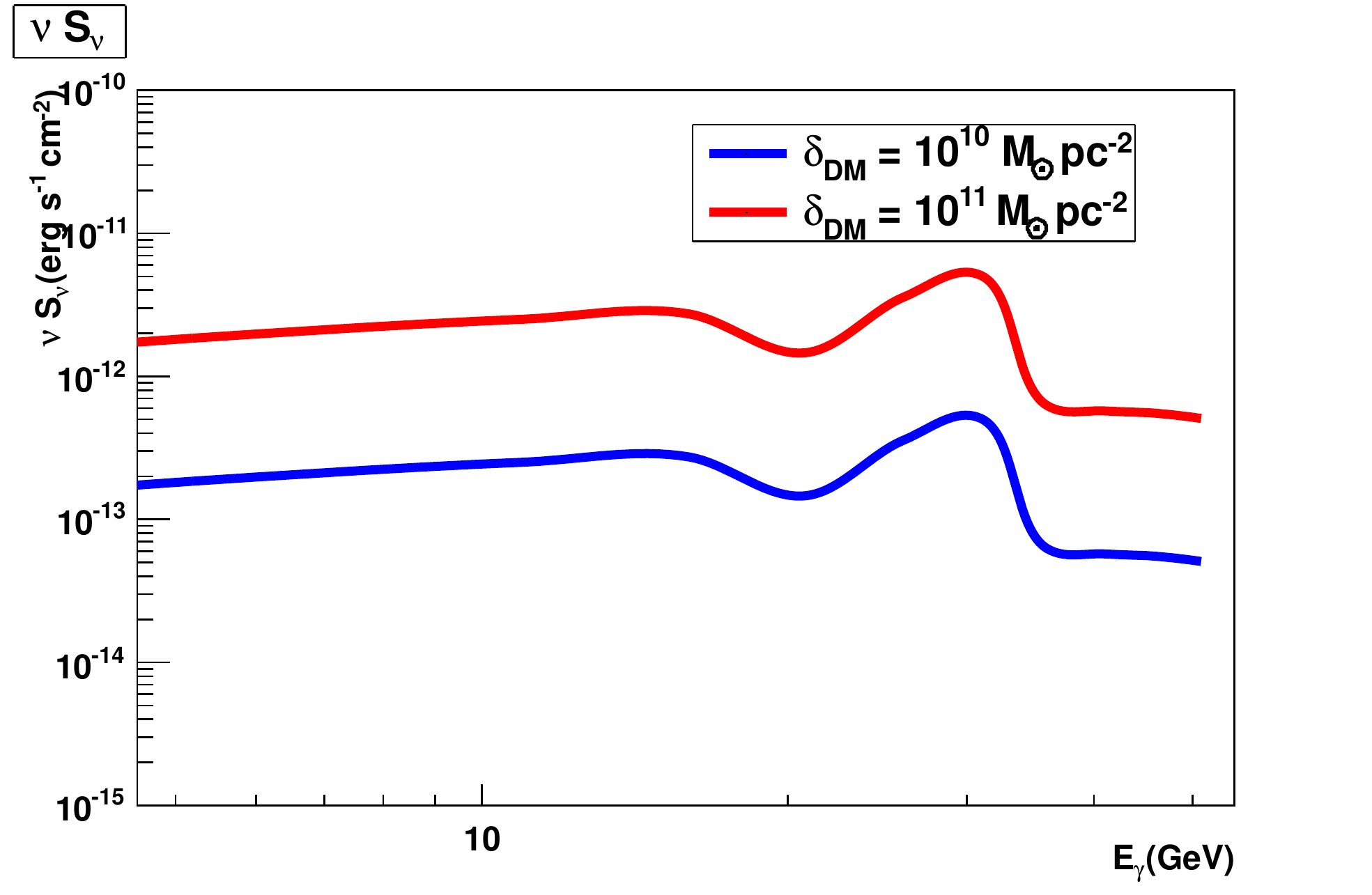}
\caption{\label{fig:photonFlux}The photon energy distributions for
$\delta_{\rm DM} = 10^{10}~M_{\odot}~{\rm pc}^{-2}$ (lower curve) and
$\delta_{\rm DM} = 10^{11}~M_{\odot}~{\rm pc}^{-2}$ (upper curve)
exhibiting a drop-off around $E_{\gamma} \sim 35$~GeV.}
\end{figure}

Putting the astrophysical inputs together with the cross section, we have:
\bea
\frac{d \Phi_{\gamma}}{d E_{\gamma}} & = &
\biggl(\delta_{\rm DM} \biggr) \times \int  ~ dE_e ~
\biggl(\frac{1}{D^2} \frac{d \Phi_{e}^{\mbox{\tiny{AGN}}}}{d E_e} \biggr) \times
\biggl( \frac{1}{m_{\tilde{\chi}}} \frac{d^2 \sigma}
{d E_{\gamma} d \Omega_{\gamma}}\biggr) .
\eea
where the three factors $\delta_{DM}$,
${d \Phi_{e}^{\mbox{\tiny{AGN}}}}/{d E_e}$ and
${d^2 \sigma}/{d E_{\gamma} d \Omega_{\gamma}}$
have been discussed in Sections~\ref{subsec:DMdist}, \ref{subsec:AGNJet}
and \ref{sec:xsec}, respectively.  We adopt two choices of
$\delta_{\rm DM} = 10^{10}~M_{\odot}~{\rm pc}^{-2}$ and
$\delta_{\rm DM} = 10^{11}~M_{\odot}~{\rm pc}^{-2}$ and to begin with,
the benchmark point with a bino neutralino of mass
$m_{\tilde{\chi}} = 60$~GeV and degenerate left- and right-handed
sleptons of masses $m_{\tilde{e}} = 100$~GeV.
In Figure~\ref{fig:photonFlux}, we plot the distribution of photon energies
for the benchmark point, for both values of $\delta_{\rm DM}$ discussed above.
We note the drop at energies around $E_\gamma \sim 35$~GeV, where,
as discussed above, the required
photon energy is so large that it is no longer produced efficiently from on-shell
intermediate selectrons.

Using the existing energy sprectrum data from Fermi LAT observations of
Centaurus A, we can put bounds on the coannihilation region of SUSY parameter
space.  Ideally, one could do a shape-based analysis, as advocated in
Ref.~\cite{Gorchtein:2010xa}, looking for a
significant drop in the energy spectrum, which occurs
as one demands the photon energy be high enough that kinematics forbid it
from arising from an on-shell selectron decay.  However, with the limited
statistics of the current data, it is more feasible to consider a counting experiment
analysis in each energy bin.
We
place $95\%$ confidence limits on regions of MSSM parameter space where a
contribution larger than the expected $95\%$ fluctuations
(based on the error bars) in the spectrum is
predicted.  We use the quoted uncertainties on the Fermi measurement points,
\bea
\sqrt{B} & \sim & 2.5 \times 10^{-12}~{\rm erg}~{\rm s}^{-1}\, {\rm cm}^{-2}~,
\eea
roughly independent of energy \cite{:2010fk}.
These bounds are
subject to our assumptions concerning $\delta_{\rm DM}$
and the beam composition and energy spectrum.

\begin{figure}[!t]
\includegraphics[width=8.00cm,angle=0]{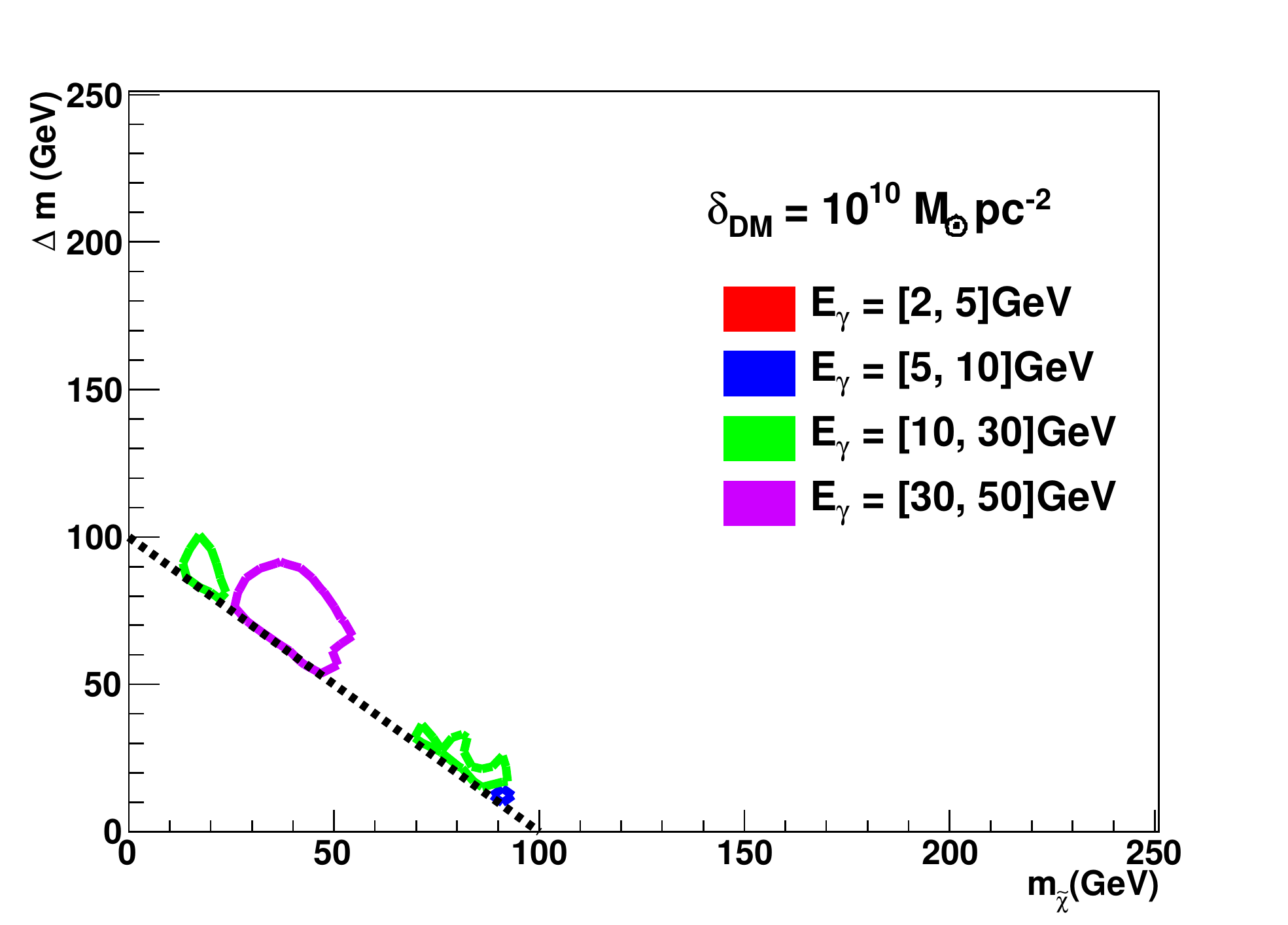}
\includegraphics[width=8.00cm,angle=0]{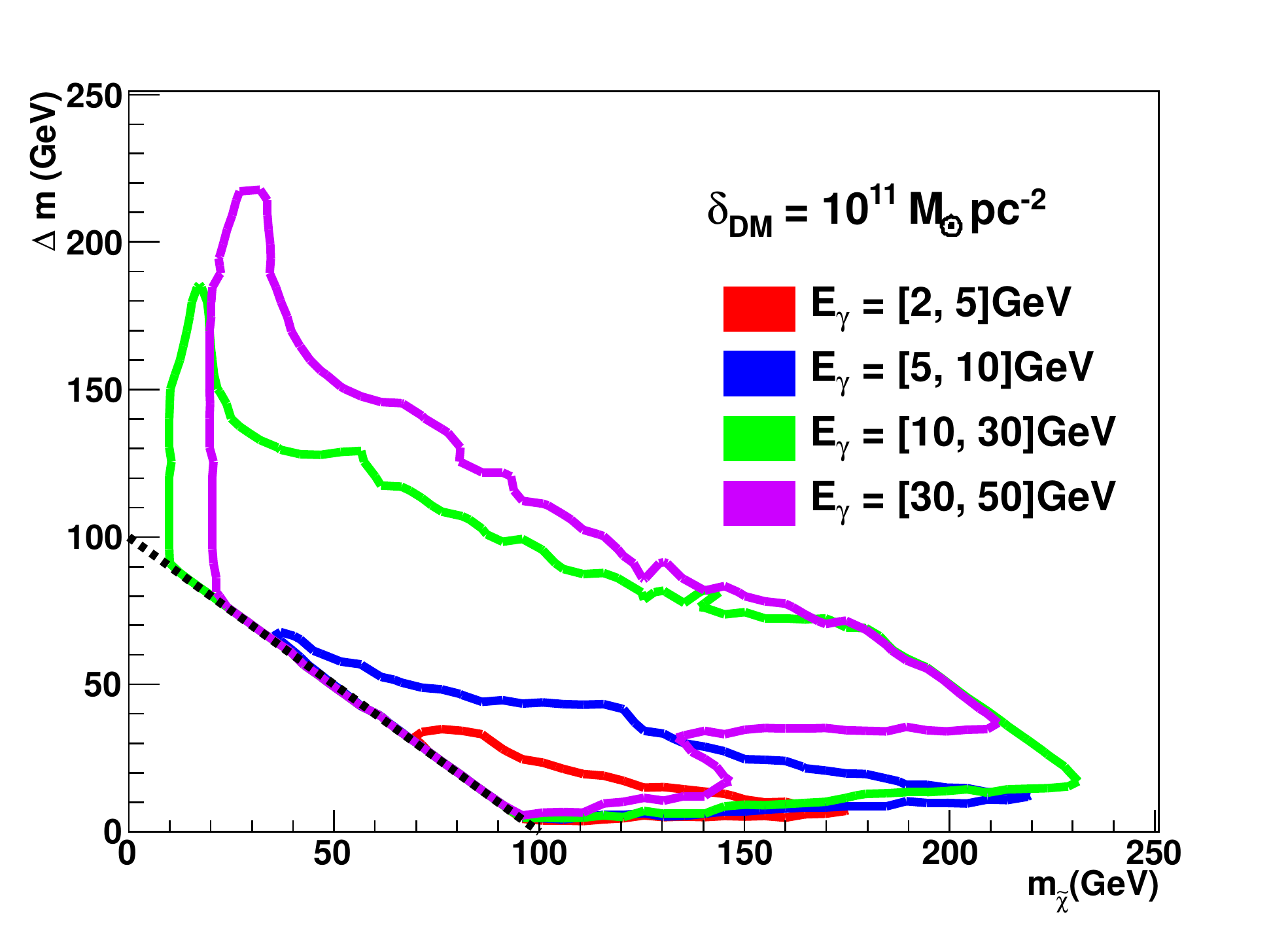}
\caption{\label{fig:contour} Contours of $95\%$ CL constraint
on the $m_{\tilde{\chi}} - \Delta {m}$ plane based on one year of observation of
Centaurus A by the Fermi LAT.  The colored regions indicate constraints
from each energy bin reported by Fermi, and the black dashed line
indicates $m_{\tilde{e}} = 100$~GeV.  The two panels assume different
$\delta_{\rm DM}$, as indicated.}
\end{figure}

In Figure~\ref{fig:contour}, we plot the $95\%$ CL bounds in the plane of
$m_{\tilde{\chi}}$ and $\Delta m$, where
\bea
\Delta m & \equiv & m_{\tilde{e}} - m_{\tilde{\chi}}
\eea
is the mass splitting between the neutralino and selectron.
The left panel corresponds to
$\delta_{\rm DM} = 10^{10}~M_{\odot}~{\rm pc}^{-2}$, and the right to
$\delta_{\rm DM} = 10^{11}~M_{\odot}~{\rm pc}^{-2}$.  These values
correspond to a neutralino annihilation cross section of
$\langle \sigma v \rangle \sim 10^{-30}$~cm$^{3}$~s$^{-1}$ (appropriate
for a coannihilation scenario) and a black hole lifetime of
$t_{\rm BH} \sim 10^8$ or $10^{10}$ years, respectively.
We continue to
assume a bino neutralino and selectrons which are degenerate in mass.  Since
the optimal energy bin for the search varies with $m_{\tilde{\chi}}$ and $m_{\tilde{e}}$,
we derive the ruled out region from each energy
bin independently, as indicated on the figure.  We find that
the largest resolving power comes from the highest energy bins, but nevertheless
the lower energy bins provide interesting constraints, and
a combined shape analysis would probably do a little better than treating
each bin independently.  The black dashed line
on each panel indicates the line of $m_{\tilde{e}} \geq 100$~GeV, roughly the
bound from LEP-II null searches \cite{Nakamura:2010zzi}.  There are no current
LHC bounds on slepton masses and what they turn out to be will ultimately
depend on the LHC performance.  For comparison, the theoretical study in
Ref. \cite{Andreev:2004qq} concluded that the LHC at $\sqrt{s} = 14$~TeV
and with $\sim 100$~fb$^{-1}$ has sensitivity for slepton masses around 100 GeV,
provided the mass splitting is larger than about $\Delta m = 20$ GeV.

\section{Conclusions and Outlook}
\label{sec:outlook}

Active galactic nuclei are among the most energetic natural accelerators in
the Universe.  They are also located in regions rich with dark matter, and provide
a great opportunity to study the high energy interactions of WIMPs with
SM particles.  In this paper, we have estimated the regions of MSSM parameter
space which can be probed from Fermi observations.  We find that there
is greatest sensitivity for models with low annihilation cross sections and nearly
degenerate sleptons -- as occurs in coannihilation regions of the MSSM, where
sleptons are active during freeze-out.  Indeed, sleptons are very challenging to
discover at the LHC under ideal circumstances, and almost impossible when
nearly degenerate with the lightest supersymmetry particle.  Observations of
the gamma rays from AGNs can help cover an interesting and important gap
in LHC coverage of MSSM parameter space.
Analogously, one can study the scattering process between dark matter and AGN jets consisting of protons,
although the rates at the highest energies will be suppressed by the parton distribution functions; we leave
this direction for future work.

Ultimately, the limiting factors in utilizing these naturally occurring accelerators as probes of the property of dark matter are the uncertainties in the properties of the AGNs themselves and the dark matter environment around them.  These uncertainties can be mitigated by further observations, to help pin down the underlying properties of the jet, and observation of the dynamics of the surrounding stars and gas to better constrain the gravitational dynamics. The MSSM signal considered here also leads to a distinctive feature in the gamma ray spectrum, which could be exploited with more statistics to minimize the dependence on the unknown background emission. Ultimately, if the particle properties of dark matter can be pinned down more precisely using data from
the LHC or direct or indirect dark matter detection experiments, one could even turn the process around and use collisions between jets and the WIMPs to learn about the astrophysics of these fascinating systems.

 \section{Acknowledgements}
T. Tait is glad to acknowledge conversations
with L.~Costamente,  S.~Profumo, and L.~Ubaldi,
and the hospitality of the SLAC theory group, for their generosity during his many visits.
The work of AR and JH is supported in part by NSF grants PHY-0653656 and PHY-0709742. The work of TMPT is
supported in part by NSF grant PHY-0970171.

\appendix
\section{Matrix Elements for $e^- \tilde{\chi} \rightarrow e^-
\tilde{\chi} \gamma$}
\label{app:amp}

In this appendix, we summarize the full expression for the amplitude squared
for the process $e^- \tilde{\chi} \rightarrow e^- \tilde{\chi} \gamma$.
We begin with some notations,
\bea
	\alpha_{\mbox{\tiny{eff}}} & = & 4 \pi \alpha_{\mbox{\tiny{EM}}}(a_{\mbox{\tiny{L}}}^4 + a_{\mbox{\tiny{R}}}^4) \; ; \nonumber \\
	s & = & (p_1 + p_2)^2 \; ; \nonumber \\
	s^{\prime} & = & (p_1^{\prime} + p_2^{\prime})^2 \; ; \nonumber \\
	u & = & (p_1 - p_2^{\prime})^2 \; ; \nonumber \\
	u^{\prime} & = & (p_2 - p_1^{\prime})^2 \; ; \nonumber \\
	E_{1s} & = & (s + m_{\tilde{\chi}}^2 - m_{e}^2)/(2\sqrt{s}) \; ; \nonumber \\
	E_{2s} & = & \sqrt{s} - E_{1s} \; ; \nonumber \\
	E_{1s^{\prime}} & = & (s^{\prime} + m_{\tilde{\chi}}^2 - m_{e}^2)/(2\sqrt{s^{\prime}}) \; ; \nonumber \\
	E_{2s^{\prime}} & = & \sqrt{s^{\prime}} - E_{1s^{\prime}} \; ; \nonumber \\
		r_{s} & = & \sqrt{(s - (m_{e} + m_{\tilde{\chi}})^2)(s - (m_{\tilde{\chi}} - m_{e})^2)}/(2\sqrt{s}) \; ; \nonumber \\
	r_{s^{\prime}} & = & \sqrt{(s^{\prime} - (m_{e} + m_{\tilde{\chi}})^2)(s^{\prime} - (m_{\tilde{\chi}} - m_{e})^2)}/(2\sqrt{s^{\prime}}) \; ; \nonumber \\
	\Gamma_1 & = & ((a_{\mbox{\tiny{L}}}^2 + a_{\mbox{\tiny{R}}}^2)(E_{1s}E_{2s} + r_{s}^2)r_{s})/(4 \pi m_{\tilde{e}}\sqrt{s}); \nonumber \\
	\Gamma_2 & = & ((a_{\mbox{\tiny{L}}}^2 + a_{\mbox{\tiny{R}}}^2)(E_{1s^{\prime}}E_{2s^{\prime}} + r_{s^{\prime}}^2)r_{s^{\prime}})/(4 \pi m_{\tilde{e}}\sqrt{s^{\prime}}) \; ; \nonumber \\
		\Sigma_{s} & = & s - m_{\tilde{e}}^2 - i\,  m_{\tilde{e}}\Gamma_1 \; ; \nonumber \\
	\Sigma_{s^{\prime}} & = & s^{\prime} - m_{\tilde{e}}^2 - i\, m_{\tilde{e}}\Gamma_2 \; ; \nonumber \\
	\Sigma_{u} & = & u - m_{\tilde{e}}^2 \; ; \nonumber \\
	\Sigma_{u^{\prime}} & = & u^{\prime} - m_{\tilde{e}}^2 \; ; \nonumber \\
		f_3 & = & 1 + i (m_{\tilde{e}}(\Gamma_1 - \Gamma_2))/(2((p_1 + p_2) \cdot r)) \; ; \nonumber \\
	|\Sigma_{s}|^2 & = &
	(s - m_{\tilde{e}}^2)^2 + (m_{\tilde{e}}~\Gamma_1)^2 \; ; \nonumber \\
	|\Sigma_{s^{\prime}}|^2 & = &
	(s^{\prime} - m_{\tilde{e}}^2)^2 + (m_{\tilde{e}}~\Gamma_2)^2 \; .
\eea
Here $p_1 \, , p_1^{\prime} \, , p_2 \, , p_2^{\prime} \, , r$ are the four-vectors for
the momenta of the incoming and outgoing dark matter particles (neutralino),
incoming and outgoing electrons and photon.
$E_{1s}$, $E_{2s}$, $E_{1s^{\prime}}$, and
$E_{2s^{\prime}}$ are the energies of the
initial and final neutralino and electron
in the lab frame where the initial neutralino is at rest.
The $s$ and $s^{\prime}$ subscripts correspond to the $s$ and $s^{\prime}$
diagrams, respectively. $\Gamma_i$ are the momentum dependent decay
widths and $\Sigma_{s}$, $\Sigma_{s^{\prime}}$ and
$\Sigma_{u}$, $\Sigma_{u^{\prime}}$ are the propagators in the
$s$- and $u$-channels. $f_3$ is the modified vertex factor in amplitude $M_3$
which insures gauge invariance, see appendix \ref{app:gauInv}.
The squared amplitudes, including the interference terms, are:
	\begin{equation}
	\begin{array}{lll}
	|M_1|^2 & = &  [\alpha_{\mbox{\tiny{eff}}}(p_1 \cdot r)(p_1^{\prime} \cdot p_2^{\prime})]/[|\Sigma_{s^{\prime}}|^2(p_2 \cdot r)] \; ; \\
	|M_2|^2 & = & [\alpha_{\mbox{\tiny{eff}}}(p_1^{\prime} \cdot r)(p_1 \cdot p_2)]/[|\Sigma_{s}|^2((p_2^{\prime} \cdot r) - \frac{m_{e}^2}{2})] \; ; \\
	|M_3|^2 & = & [-4\alpha_{\mbox{\tiny{eff}}}(p_1 \cdot p_2)(p_1^{\prime} \cdot p_2^{\prime})(2(p_1 \cdot p_2) - (p_1 \cdot r) - (p_2 \cdot r) + m_{\tilde{\chi}}^2)|f_3|^2]/[|\Sigma_{s}|^2|\Sigma_{s^{\prime}}|^2] \; ; \\	
	2Re(M_1M_2^{\dagger}) & = & Re\biggl[\frac{-\alpha_{\mbox{\tiny{eff}}}}{\Sigma_{s^{\prime}}\Sigma_{s}^{*}(p_2 \cdot r)((p_2^{\prime} \cdot r) - m_{e}^2/2)} \biggl((p_1 \cdot p_2^{\prime})(p_1 \cdot r)(p_2 \cdot r) - (p_1 \cdot p_2^{\prime})(p_1 \cdot r)(p_1 \cdot p_2) \\
	 & & {} + (p_1 \cdot p_2^{\prime})(p_2 \cdot r)^2 - 3(p_1 \cdot p_2^{\prime})(p_2 \cdot r)(p_1 \cdot p_2) + 2(p_1 \cdot p_2^{\prime})(p_1 \cdot p_2)^2 \\
	 & & {} - (p_2 \cdot p_2^{\prime})(p_1 \cdot r)^2 - (p_2 \cdot p_2^{\prime})(p_1 \cdot r)(p_2 \cdot r) + (p_2 \cdot p_2^{\prime})(p_1 \cdot r)(p_1 \cdot p_2) \\
	 & & {} - (p_2 \cdot p_2^{\prime})(p_2 \cdot r)(p_1 \cdot p_2) - (p_1 \cdot r)^2(p_1 \cdot p_2) - 2(p_1 \cdot r)(p_2 \cdot r)(p_1 \cdot p_2) \\
	 & & {} + 3(p_1 \cdot r)(p_1 \cdot p_2)^2 - (p_2 \cdot r)^2(p_1 \cdot p_2) + 3(p_2 \cdot r)(p_1 \cdot p_2)^2 - 2(p_1 \cdot p_2)^3 \biggr) \biggr] \; ; \\
	2Re(M_1M_3^{\dagger}) & = & Re\biggl[\frac{2\alpha_{\mbox{\tiny{eff}}}f_3}{|\Sigma_{s^{\prime}}|^2\Sigma_{s}^{*}(p_2 \cdot r)}
	\biggl(2(p_1 \cdot r)^2(p_2 \cdot p_1) + 3(p_1 \cdot r)(p_2 \cdot r)(p_1 \cdot p_2) \\
	& & {} - (p_1 \cdot r)(p_2 \cdot r)m_{\tilde{\chi}}^2 - 4(p_1 \cdot r)(p_2 \cdot p_1)^2 + (p_2 \cdot r)^2(p_1 \cdot p_2) \\
	& & {} - (p_2 \cdot r)^2m_{\tilde{\chi}}^2 - 3(p_2 \cdot r)(p_2 \cdot p_1)^2 + (p_2 \cdot r)(p_1 \cdot p_2)m_{\tilde{\chi}}^2 + 2(p_2 \cdot p_1)^3 \biggr)\biggr] \; ; \\
	2Re(M_2M_3^{\dagger}) & = & Re\biggl[\frac{-2\alpha_{\mbox{\tiny{eff}}}f_3}{|\Sigma_{s}|^2\Sigma_{s^{\prime}}^{*}((p_2^{\prime} \cdot r) - m_{e}^2/2)}
	\biggl((p_1 \cdot p_2^{\prime})(p_1 \cdot r)(p_1 \cdot p_2)+(p_1 \cdot p_2^{\prime})(p_2 \cdot r)(p_1 \cdot p_2) \\
	& & {} - (p_1 \cdot p_2^{\prime})(p_2 \cdot p_1)^2 - (p_1 \cdot p_2^{\prime})(p_1 \cdot p_2)m_{\tilde{\chi}}^2 + (p_2 \cdot p_2^{\prime})(p_1 \cdot r)(p_1 \cdot p_2) \\
	& & {} + (p_2 \cdot p_2^{\prime})(p_2 \cdot r)(p_1 \cdot p_2) - (p_2 \cdot p_2^{\prime})(p_1 \cdot p_2)^2 - (p_2 \cdot p_2^{\prime})(p_2 \cdot p_1)m_{\tilde{\chi}}^2 \\
	& & {} + (p_1 \cdot r)^2(p_1 \cdot p_2) + 2(p_1 \cdot r)(p_2 \cdot r)(p_1 \cdot p_2) - 4(p_1 \cdot r)(p_1 \cdot p_2)^2 \\
	& & {} - (p_1 \cdot r)(p_1 \cdot p_2)m_{\tilde{\chi}}^2 + (p_2 \cdot r)^2(p_1 \cdot p_2) - 4(p_2 \cdot r)(p_1 \cdot p_2)^2 \\
	& & {} - (p_2 \cdot r)(p_1 \cdot p_2)m_{\tilde{\chi}}^2 + 3(p_1 \cdot p_2)^3 + (p_1 \cdot p_2)^2m_{\tilde{\chi}}^2 \biggr) \biggr] \; ; 	
\end{array}
\end{equation}	

\begin{equation}
\begin{array}{lll}
	|M_4|^2 & = & [\alpha_{\mbox{\tiny{eff}}}(p_1^{\prime} \cdot r)(p_1 \cdot p_2^{\prime})]/[\Sigma_{u}^2(p_2 \cdot r)] \; ; \\
	|M_5|^2 & = & [\alpha_{\mbox{\tiny{eff}}}(p_1 \cdot r)(p_1^{\prime} \cdot p_2)]/[\Sigma_{u^{\prime}}^2(p_2^{\prime} \cdot r - m_{e}^2/2)] \; ; \\
	|M_6|^{2} & = & [4\alpha_{\mbox{\tiny{eff}}}(p_1 \cdot p_2^{\prime})(p_2 \cdot p_1^{\prime})((p_2 \cdot p_1^{\prime}) + (p_1 \cdot p_2^{\prime}) - m_{\tilde{\chi}}^2)]/[\Sigma_{u}^2\Sigma_{u^{\prime}}^2] \\
	2Re(M_4M_5^{\dagger}) & = & Re\biggl[\frac{\alpha_{\mbox{\tiny{eff}}}}{\Sigma_{u}\Sigma_{u^{\prime}}(p_2 \cdot r)((p_2^{\prime} \cdot r) - m_{e}^2/2)}
	\biggl(2(p_1 \cdot p_2^{\prime})^2(p_2 \cdot p_2^{\prime}) + (p_1 \cdot p_2^{\prime})^2(p_2 \cdot r) \\
	& & {} - (p_1 \cdot p_2^{\prime})^2(p_1 \cdot p_2) + (p_1 \cdot p_2^{\prime})(p_2 \cdot p_2^{\prime})(p_1 \cdot r) + (p_2 \cdot p_1)^3 \\
	& & {} - (p_1 \cdot p_2^{\prime})(p_2 \cdot p_2^{\prime})(p_2 \cdot r) - (p_1 \cdot p_2^{\prime})(p_1 \cdot r)(p_2 \cdot p_1) - (p_1 \cdot p_2^{\prime})(p_2 \cdot r)^2 \\
	& & {} + (p_1 \cdot p_2^{\prime})(p_2 \cdot r)(p_1 \cdot p_2) - (p_2 \cdot p_2^{\prime})^2(p_1 \cdot r) + (p_2 \cdot p_2^{\prime})^2(p_1 \cdot p_2) \\
	& & {} - (p_2 \cdot p_2^{\prime})(p_1 \cdot r)(p_2 \cdot r) + 2(p_2 \cdot p_2^{\prime})(p_1 \cdot r)(p_2 \cdot p_1) + 2(p_2 \cdot p_2^{\prime})(p_2 \cdot r)(p_1 \cdot p_2) \\
	& & {} - 2(p_2 \cdot p_2^{\prime})(p_2 \cdot p_1)^2 + (p_1 \cdot r)(p_2 \cdot r)(p_1 \cdot p_2) - (p_1 \cdot r)(p_1 \cdot p_2)^2\\
	& & {} + (p_2 \cdot r)^2(p_1 \cdot p_2) - 2(p_2 \cdot r)(p_1 \cdot p_2)^2 \biggr) \biggr]\; ; \\
	2Re(M_4M_6^{\dagger}) & = & Re \biggl[\frac{2\alpha_{\mbox{\tiny{eff}}}}{\Sigma_{u}^2\Sigma_{u^{\prime}}(p_2 \cdot r)}
	\biggl(2(p_1 \cdot p_2^{\prime})^2(p_2 \cdot p_2^{\prime}) + 2(p_1 \cdot p_2^{\prime})^2(p_2 \cdot r) \\
	& & {} - 2(p_1 \cdot p_2^{\prime})^2(p_1 \cdot p_2) - (p_1 \cdot p_2^{\prime})(p_2 \cdot p_2^{\prime})(p_2 \cdot r) - (p_1 \cdot p_2^{\prime})(p_2 \cdot r)^2 \\
	& & {} + (p_1 \cdot p_2^{\prime})(p_2 \cdot r)(p_1 \cdot p_2) - (p_1 \cdot p_2^{\prime})(p_2 \cdot r)m_{\tilde{\chi}}^2 \biggr) \biggr] \; ; \\
	2Re(M_5M_6^{\dagger}) & = & Re \biggl[\frac{-2\alpha_{\mbox{\tiny{eff}}}}{\Sigma_{u}^2\Sigma_{u^{\prime}}^2((p_2^{\prime} \cdot r) - m_{e}^2/2)}
	\biggl((p_1 \cdot p_2^{\prime})^2(p_2 \cdot p_2^{\prime}) + (p_1 \cdot p_2^{\prime})^2(p_2 \cdot r) \\
	& & {} - (p_1 \cdot p_2^{\prime})^2(p_1 \cdot p_2) - (p_1 \cdot p_2^{\prime})(p_2 \cdot p_2^{\prime})^2 + (p_1 \cdot p_2^{\prime})(p_2 \cdot p_2^{\prime})(p_1 \cdot r) \\
	& & {} - 2(p_1 \cdot p_2^{\prime})(p_2 \cdot p_2^{\prime})(p_2 \cdot r) +2(p_1 \cdot p_2^{\prime})(p_2 \cdot p_2^{\prime})(p_1 \cdot p_2) - (p_1 \cdot p_2^{\prime})(p_2 \cdot p_2^{\prime})m_{\tilde{\chi}}^2 \\
	& & {} + (p_1 \cdot p_2^{\prime})(p_1 \cdot r)(p_2 \cdot r) - (p_1 \cdot p_2^{\prime})(p_1 \cdot r)(p_1 \cdot p_2) - (p_1 \cdot p_2^{\prime})(p_2 \cdot r)^2 \\
	& & {} + 2(p_1 \cdot p_2^{\prime})(p_2 \cdot r)(p_1 \cdot p_2) - (p_1 \cdot p_2^{\prime})(p_2 \cdot r)m_{\tilde{\chi}}^2 - (p_1 \cdot p_2^{\prime})(p_1 \cdot p_2)^2 \\
	& & {} + (p_1 \cdot p_2^{\prime})(p_1 \cdot p_2)m_{\tilde{\chi}}^2 - (p_2 \cdot p_2^{\prime})^2m_{\tilde{\chi}}^2 - (p_2 \cdot p_2^{\prime})(p_1 \cdot r)m_{\tilde{\chi}}^2 \\
	& & {} - 2(p_2 \cdot p_2^{\prime})(p_2 \cdot r)m_{\tilde{\chi}}^2 + 2(p_2 \cdot p_2^{\prime})(p_1 \cdot p_2)m_{\tilde{\chi}}^2 - (p_1 \cdot r)(p_2 \cdot r)m_{\tilde{\chi}}^2 \\
	& & {} + (p_1 \cdot r)(p_1 \cdot p_2)m_{\tilde{\chi}}^2 - (p_2 \cdot r)^2m_{\tilde{\chi}}^2 + 2(p_2 \cdot r)(p_1 \cdot p_2)m_{\tilde{\chi}}^2 - (p_1 \cdot p_2)^2m_{\tilde{\chi}}^2 \biggr) \biggr] \; ;
\end{array}	
\end{equation}	

\begin{equation}
\begin{array}{lll}
	2Re(M_1M_4^{\dagger})  & = & Re \biggl[\frac{\alpha_{\mbox{\tiny{eff}}} \, m_{\tilde{\chi}}^2}{\Sigma_{s^{\prime}}\Sigma_{u}(p_2 \cdot r)}\biggl((p_1 \cdot p_2^{\prime}) + (p_2 \cdot p_2^{\prime}) + (p_1 \cdot r) + (p_2 \cdot r) - (p_1 \cdot p_2)\biggr)\biggr] \; ; \\
	2Re(M_1M_5^{\dagger}) & = & Re \biggl[\frac{-\alpha_{\mbox{\tiny{eff}}} \, m_{\tilde{\chi}}^2}{\Sigma_{s^{\prime}}\Sigma_{u^{\prime}}(p_2 \cdot r)((p_2^{\prime} \cdot r) - m_{e}^2/2)}\biggl((p_1 \cdot p_2^{\prime})(p_2 \cdot p_2^{\prime}) + (p_2 \cdot p_2^{\prime})(p_1 \cdot r) - (p_2 \cdot p_2^{\prime})(p_1 \cdot p_2)\biggr) \biggr] \; ; \\
	2Re(M_1M_6^{\dagger}) & = & Re \biggl[\frac{-\alpha_{\mbox{\tiny{eff}}}\, m_{\tilde{\chi}}^2}{\Sigma_{s^{\prime}}\Sigma_{u^{\prime}}\Sigma_{u}(p_2 \cdot r)}
	\biggl(2(p_1 \cdot p_2^{\prime})(p_2 \cdot p_2^{\prime}) + (p_1 \cdot p_2^{\prime})(p_2 \cdot r) \\
	& & {} - (p_1 \cdot p_2^{\prime})(p_1 \cdot p_2) + (p_2 \cdot p_2^{\prime})(p_1 \cdot r) + (p_2 \cdot p_2^{\prime})(p_2 \cdot r) \\
	& & {} - (p_2 \cdot p_2^{\prime})(p_1 \cdot p_2) - (p_1 \cdot r)(p_1 \cdot p_2) - (p_2 \cdot r)(p_1 \cdot p_2) + (p_1 \cdot p_2)^2 \biggr) \biggr] \; ; \\
	2Re(M_4M_2^{\dagger}) & = & Re \biggl[\frac{-\alpha_{\mbox{\tiny{eff}}} \, m_{\tilde{\chi}}^2}{\Sigma_{u}\Sigma_{s}^{*}(p_2 \cdot r)((p_2^{\prime} \cdot r) - m_{e}^2/2)}\biggl((p_1 \cdot p_2^{\prime})(p_2 \cdot p_2^{\prime}) + (p_2 \cdot p_2^{\prime})(p_1 \cdot r) - (p_2 \cdot p_2^{\prime})(p_1 \cdot p_2)\biggr)\biggr] \; ; \\
	2Re(M_2M_5^{\dagger}) & = & Re\biggl[[\alpha_{\mbox{\tiny{eff}}} \, m_{\tilde{\chi}}^2 (p_2 \cdot r)]/[\Sigma_{s}\Sigma_{u^{\prime}}((p_2^{\prime} \cdot r) - m_{e}^2/2)]\biggr] \; ; \\
	2Re(M_2M_6^{\dagger}) & = & Re\biggl[\frac{\alpha_{\mbox{\tiny{eff}}} \, m_{\tilde{\chi}}^2}{\Sigma_{s}\Sigma_{u}\Sigma_{u^{\prime}}((p_2^{\prime} \cdot r)  - m_{e}^2/2)}
	\biggl((p_1 \cdot p_2^{\prime})(p_2 \cdot p_2^{\prime}) + (p_1 \cdot p_2^{\prime})(p_2 \cdot r) \\
	& & {} - (p_1 \cdot p_2^{\prime})(p_1 \cdot p_2) - (p_2 \cdot p_2^{\prime})^2 - (p_2 \cdot p_2^{\prime})(p_2 \cdot r) \\
	& & {} - (p_1 \cdot r)(p_1 \cdot p_2) - (p_2 \cdot r)(p_1 \cdot p_2) + (p_1 \cdot p_2)^2 \biggr) \biggr]; \\
	2Re(M_4M_3^{\dagger}) & = & Re \biggl[\frac{\alpha_{\mbox{\tiny{eff}}} \, m_{\tilde{\chi}}^2 \, f_3}{\Sigma_{u}\Sigma_{s}^{*}\Sigma_{s^{\prime}}^{*}(p_2 \cdot r)}
	\biggl((p_1 \cdot p_2^{\prime})(p_2 \cdot r) - (p_1 \cdot p_2^{\prime})(p_1 \cdot p_2) \\
	& & {} - (p_2 \cdot p_2^{\prime})(p_1 \cdot r) - (p_2 \cdot p_2^{\prime})(p_2 \cdot r) + (p_2 \cdot p_2^{\prime})(p_1 \cdot p_2) \\
	& & {} - (p_1 \cdot r)(p_1 \cdot p_2) - (p_2 \cdot r)(p_1 \cdot p_2) + (p_1 \cdot p_2)^2 \biggr) \biggr]; \\
	2Re(M_5M_3^{\dagger}) & = & Re \biggl[\frac{-\alpha_{\mbox{\tiny{eff}}} \, m_{\tilde{\chi}}^2 \, f_3}{\Sigma_{u^{\prime}}\Sigma_{s}^{*}\Sigma_{s^{\prime}}^{*}((p_2^{\prime} \cdot r) - m_{e}^2/2)}
	\biggl((p_1 \cdot p_2^{\prime})(p_2 \cdot p_2^{\prime}) + (p_1 \cdot p_2^{\prime})(p_2 \cdot r) \\
	& & {} - (p_1 \cdot p_2^{\prime})(p_1 \cdot p_2) + (p_2 \cdot p_2^{\prime})^2 + (p_2 \cdot p_2^{\prime})(p_2 \cdot r) \\
	& & {} - (p_1 \cdot r)(p_1 \cdot p_2) - (p_2 \cdot r)(p_1 \cdot p_2) + (p_1 \cdot p_2)^2 \biggr) \biggr]; \\
	2Re(M_3M_6^{\dagger}) & = & Re \biggl[\frac{-2\alpha_{\mbox{\tiny{eff}}} \, m_{\tilde{\chi}}^2 \, f_3^{*}}{\Sigma_{s}\Sigma_{s^{\prime}}\Sigma_{u}\Sigma_{u^{\prime}}}
	\biggl((p_1 \cdot p_2^{\prime})(p_2 \cdot p_2^{\prime}) + (p_2 \cdot p_2^{\prime})^2 \\
	& & {} + (p_2 \cdot p_2^{\prime})(p_1 \cdot r) - (p_2 \cdot p_2^{\prime})(p_1 \cdot p_2) - 2(p_2 \cdot p_2^{\prime})m_{\tilde{\chi}}^2\biggr)\biggr] \; .	
\end{array}
\end{equation}	

\section{Proof of Gauge Invariance}
\label{app:gauInv}
In this appendix, we verify the invariance of the amplitude under $U(1)_{\rm EM}$
gauge transformations.  We begin with the set of $u$-channel Feynman
graphs.
The amplitudes of $M_4, M_5$ and $M_6$ can be written as,
\bea
\label{eqn:M4-M6}
M_4 & = & \biggl[\overline{\tilde{\chi}}(p_1^{\prime})(a_LP_L+a_RP_R){i\over \gamma.(p_2 - r) - m_e}(-ie\gamma^\mu){e}(p_2)\biggr] \nonumber \\
& & \biggl[\bar{e}(p_2^{\prime}) (a_LP_L+a_RP_R)\chi(p_1)\biggr]\biggl[{i\over (p_2-p_1^{\prime} - r)^2-m_{\tilde{e}}^2} \biggr] \; , \nonumber \\
M_5 & = & \biggl[\overline{\tilde{\chi}}(p_1^{\prime})(a_LP_L+a_RP_R){e}(p_2)\biggr]\biggl[{i\over (p_2-p_1^{\prime})^2-m_{\tilde{e}}^2}\biggr] \nonumber \\
& & \biggl[\bar{e}(p_2^{\prime})(-ie\gamma^\mu){i\over \gamma.(p_2^{\prime} + r)-m_e} (a_LP_L+a_RP_R)\chi(p_1)\biggr] \; , \nonumber \\
M_6 & = & \biggl[\overline{\tilde{\chi}}(p_1^{\prime})(a_LP_L+a_RP_R){e}(p_2)\biggr]\biggl[\bar{e}(p_2^{\prime}) (a_LP_L+a_RP_R)\chi(p_1)\biggr] \nonumber \\
& & \biggl[{i\over (p_2-p_1^{\prime})^2-m_{\tilde{e}}^2}(-ie)(2p_2 - 2p_1^{\prime}- r)^\mu {i\over (p_2-p_1^{\prime} - r)^2-m_{\tilde{e}}^2}\biggr] \; .
\eea
Applying the Ward identity to $M_4$-$M_6$ individually gives rise to
\bea
M_4 & = &
e\biggl[\overline{\tilde{\chi}}(p_1^{\prime})(a_LP_L+a_RP_R){e}(p_2)\biggr]
\biggl[\bar{e}(p_2^{\prime}) (a_LP_L+a_RP_R)\tilde{\chi}(p_1)\biggr]
\biggl[{i\over (p_2-p_1^{\prime} - r)^2-m_{\tilde{e}}^2}\biggr] \; , \nonumber \\
M_5 & = &
-e\biggl[\overline{\tilde{\chi}}(p_1^{\prime})(a_LP_L+a_RP_R){e}(p_2)\biggr]
\biggl[\bar{e}(p_2^{\prime}) (a_LP_L+a_RP_R)\tilde{\chi}(p_1)\biggr]
\biggl[{i\over (p_2-p_1^{\prime})^2-m_{\tilde{e}}^2}\biggr] \;, \nonumber \\
M_6 & = &
e\biggl[\overline{\tilde{\chi}}(p_1^{\prime})(a_LP_L+a_RP_R){e}(p_2)\biggr]
\biggl[\bar{e}(p_2^{\prime}) (a_LP_L+a_RP_R)\tilde{\chi}(p_1) \biggr] \nonumber \\
& & \biggl[{i\over (p_2-p_1^{\prime})^2-m_{\tilde{e}}^2}- {i\over (p_2-p_1^{\prime} - r)^2-m_{\tilde{e}}^2}\biggr] \;,
\eea
and they sum to zero as expected.

\begin{figure}[t]
\includegraphics[width=12.0cm]{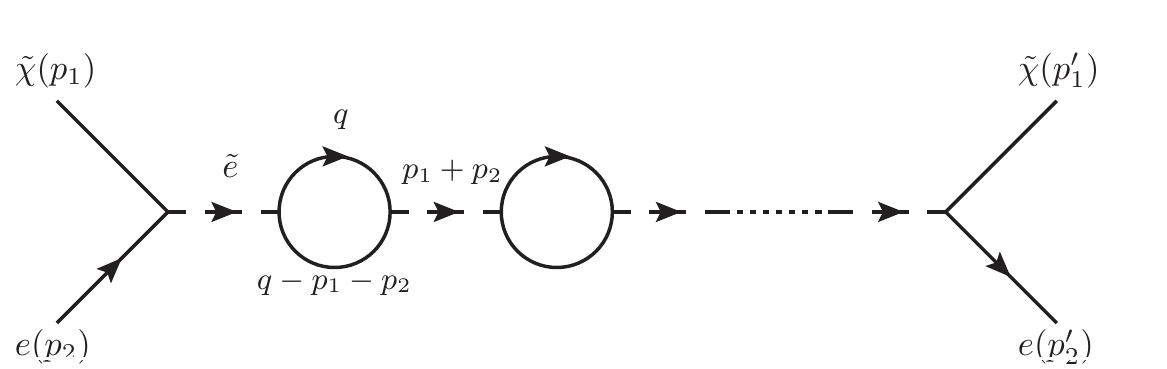}
\caption{\label{fig:bubble1} A general ``bubble" diagram
containing no photon radiation.}
\end{figure}

The set of $s$-channel diagrams is somewhat more complicated.
To begin with, we consider the process with no final state photon,
but a sum of bubble diagrams as shown in Fig.~\ref{fig:bubble1}, which
will be useful later.  The key is the imaginary part of the loop which produces the
width of the selectron.
Writting the sum of diagrams with all numbers of bubble insertions as,
\beq
{i\over p^2-m_{\tilde{e}}^2}+{i\over p^2-m_{\tilde{e}}^2}(-i\Sigma(p)){i\over p^2 - m_{\tilde{e}}^2}+...
={i\over p^2-m_{\tilde{e}}^2-\Sigma(p)}
\eeq
we have
\beq
-i\Sigma(p)=-\int {d^4q\over (2\pi)^4}
{\rm Tr}
\left[{i\over \gamma.q-m_e} \left( a_LP_L+a_R P_R \right)
{i\over \gamma.(q-p)-m_{\tilde{\chi}}} \left( a_LP_L+a_R P_R \right) \right] \; ,
\eeq
where the width is determined by
$m\Gamma=Im(\Sigma(p))$ at $p^2=m_{\tilde{e}}^2$.
The diagram with no photon radiation is further simplified to,
\beq
\biggl[\overline{\tilde{\chi}}(p_1)
\left( a_LP_L+a_R P_R \right) e(p_2)\biggr]
\biggl[{i\over p^2-m_{\tilde{e}}^2-im_{\tilde{e}}\Gamma}\biggr]
\biggl[\bar{e}(p_2^{\prime})
\left( a_LP_L+a_R P_R \right)\tilde{\chi}(p_1^{\prime})\biggr]
\eeq
with $p = p_1 + p_2$.

When the photon line is added, we get new diagrams by attaching the photon in all possible ways. The first possibility is
\begin{figure}[tbh!]
\includegraphics[width=12.0cm]{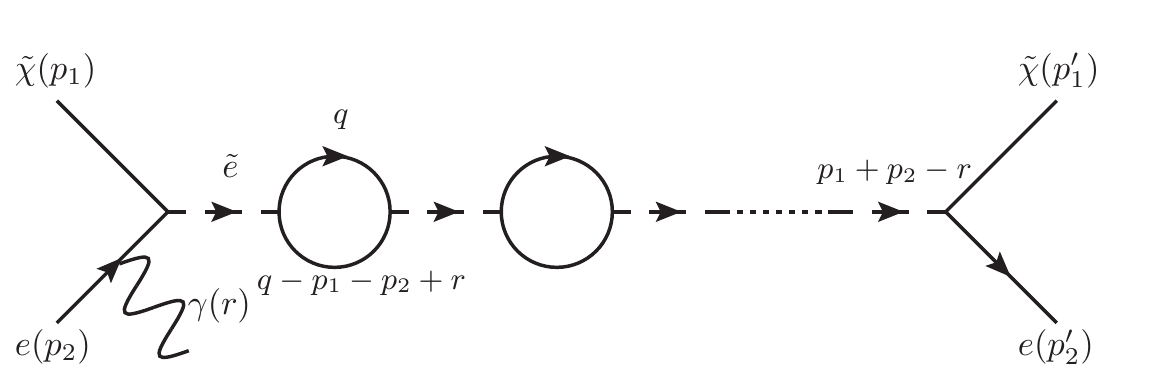}
\end{figure}
\bea
M_1 & = & \biggl[\overline{\tilde{\chi}}(p_1)(a_LP_L+a_RP_R){i\over \gamma.(p_2 - r)-m_e}(-ie)\gamma^\mu e(p_2)\biggr]\biggl[{i\over (p_1^{\prime}+p_2^{\prime})^2-m_{\tilde{e}}^2-im_{\tilde{e}}\Gamma(p_1 + p_2 - r)}\biggr] \nonumber \\
& & \biggl[\bar{e}(p_2^{\prime})(a_LP_L+a_RP_R)\tilde{\chi}(p_1^{\prime})\biggr]
\eea
We have taken the momentum of the photon (with polarization $\mu$)
to be $r$ going out from the diagram.  This graph corresponds to $M_1$.
We also have the diagram corresponding to $M_2$,
\begin{figure}[tbh!]
\includegraphics[width=12.0cm]{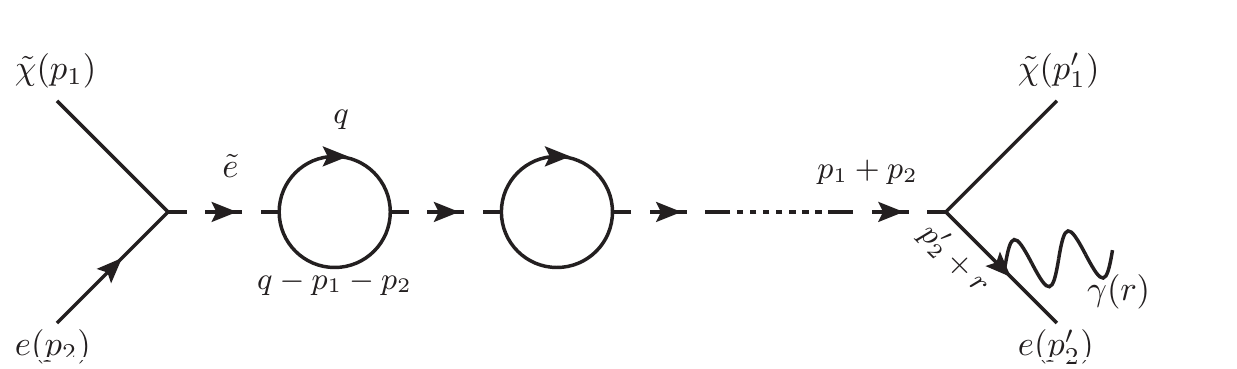}
\end{figure}
\bea
M_2 & = & \biggl[\overline{\tilde{\chi}}(p_1)(a_LP_L+a_RP_R)e(p_2)\biggr] \biggl[{i\over (p_1 + p_2)^2-m_{\tilde{e}}^2 -im_{\tilde{e}}\Gamma(p_1 + p_2)}\biggr] \nonumber \\
& & \biggl[\bar{e}(p_2^{\prime})(-ie)\gamma^\mu{i\over \gamma.(p_2^{\prime} + r)}(a_LP_L+a_RP_R)\tilde{\chi}(p_1^{\prime})\biggr]
\eea
While for $M_3$, the photon attaches to the
selectron; there are two diagrams where one is
\begin{figure}[tbh!]
\includegraphics[width=12.0cm]{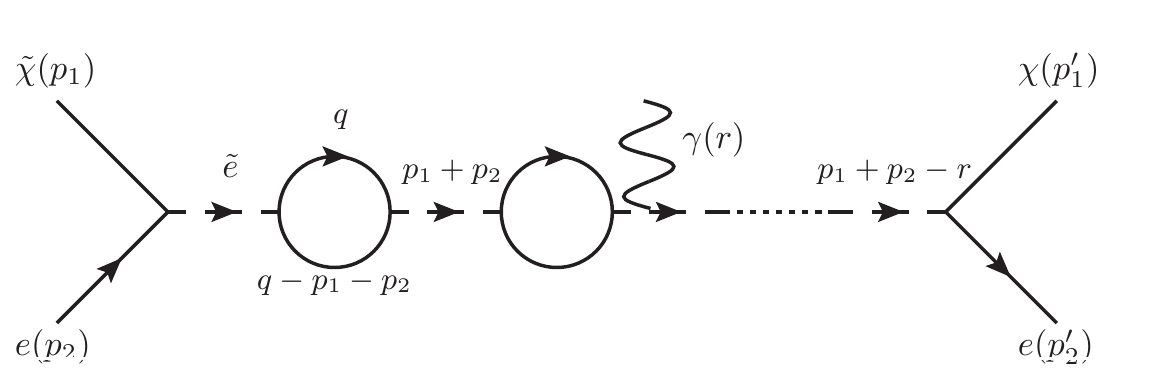}
\end{figure}
\bea
M_{3,1} & = & \biggl[\overline{\tilde{\chi}}(p_1)(a_LP_L+a_RP_R) e(p_2)\biggr]\biggl[\bar{e}(p_2^{\prime})(a_LP_L+a_RP_R)\tilde{\chi}(p_1^{\prime}) \biggr]\biggl[(-ie)(2p_2+2p_1 - r)^\mu \biggr] \nonumber \\
& & \biggl[{i\over (p_1 + p_2)^2-m_{\tilde{e}}^2-im_{\tilde{e}}\Gamma(p_1+p_2)}{i\over (p_2 + p_2 - r)^2-m_{\tilde{e}}^2-im_{\tilde{e}}\Gamma(p_1 + p_2 - r)}\biggr]
\eea
and the other one has the photon attached to an internal loop,
\begin{figure}[tbh!]
\includegraphics[width=12.0cm]{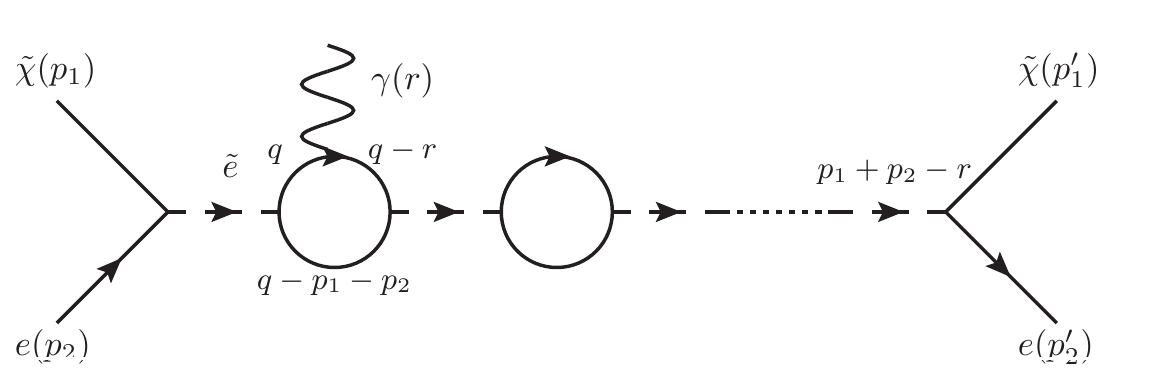}
\end{figure}
\bea
M_{3,2} & = & \biggl[\overline{\tilde{\chi}}(p_2)(a_LP_L+a_RP_R) e(p_1)\biggr] \biggl[\bar{e}(p_1^{\prime})(a_LP_L+a_RP_R)\tilde{\chi}(p_2^{\prime}) \biggr] \biggl[-i\Sigma^{\mu}(p_1 + p_2, -r)\biggr]\nonumber \\
& & \biggl[{i\over (p_1+p_2)^2-m_{\tilde{e}}^2-im_{\tilde{e}}\Gamma(p_1+p_2)} {i\over (p_1 + p_2 - r)^2-m_{\tilde{e}}^2-im_{\tilde{e}}\Gamma(p_1 + p_2 - r)} \biggr].
\eea
Now we need to extract the imaginary part of $-i\Sigma^{\mu}(p, -r)$ in this vertex,
\bea
-i\Sigma^{\mu}(p, -r) & = & \int {d^4q\over (2\pi)^4} {\rm Tr}
\biggl[{i\over \gamma.(q - r)-m_e} (-ie)\gamma^\mu {i\over \gamma.q-m_e} \nonumber \\
& & (a_LP_L+a_RP_R){i\over \gamma.(q-p)-m_{\tilde{\chi}}}(a_LP_L+a_RP_R)\biggr] \nonumber \\
& = & -\int {d^4q\over (2\pi)^4} {\rm Tr}
\biggl[{i\over \gamma.(q - r)-m_e}(-ie) \gamma.(-r) {i\over \gamma.q-m_e} \nonumber \\
& & (a_LP_L+a_RP_R){i\over \gamma.(q-p)-m_{\tilde{\chi}}}(a_LP_L+a_RP_R)\biggr] \nonumber \\
& = & -(-e)\int {d^4q\over (2\pi)^4} {\rm Tr}
\biggl[({i\over \gamma.(q - r)-m_e} - {i\over \gamma.q-m_e})\nonumber \\
& & (a_LP_L+a_RP_R){i\over \gamma.(q-p)-m_{\tilde{\chi}}}(a_LP_L+a_RP_R)\biggr] \nonumber \\
& = & (-e)(-i\Sigma(p - r)+i\Sigma(p)) \;.
\eea
We can write generally,
$-i\Sigma^{\mu}(p,-r) =A~(p+p')^\mu +B ~r^\mu$. The equation above tells us that
\beq
A(p+p').(-r) = (-e)(-i\Sigma(p - r)+i\Sigma(p))
\\
A={(-e)(-i\Sigma(p - r)+i\Sigma(p))\over -2p \cdot r} \;.
\eeq
The coefficient $B$ is irrelevant since $r \cdot \epsilon=0$, and the vertex is
effectively
$-i\Sigma^{\mu}(p,-r)=(-ie)(p+p')^\mu
{(-\Sigma(p - r)+\Sigma(p))\over -2p \cdot r}$
which can be thought of as a correction to the tree level vertex.
Since we are only interested in the imaginary piece of the correction,
\beq
(-ie)(p+p')^\mu{(-\Gamma(p - r)+\Gamma(p))\over -2p \cdot r},
\eeq
we can combine the last two diagrams, arriving at the expression,
\bea
M_3 & = & \biggl[\overline{\tilde{\chi}}(p_1)(a_LP_L+a_RP_R) e(p_2)\biggr] \biggl[\bar{e}(p_2^{\prime})(a_LP_L+a_RP_R)\tilde{\chi}(p_1^{\prime}) \biggr] \biggl[1+im_{\tilde{e}}{\Gamma(p_1+p_2)-\Gamma(p_1+p_2-r)\over 2(p_1+p_2)\cdot(-r)}\biggr] \nonumber\\
& & {i\over (p_1 + p_2)^2-m_{\tilde{e}}^2-im_{\tilde{e}}\Gamma(p_1+p_2)}(-ie)(p_1+p_2-r)^\mu {i\over (p_1+p_2-r)^2-m_{\tilde{e}}^2-im_{\tilde{e}}\Gamma(p_1+p_2-r)} \nonumber \\
& = & \biggl[\overline{\tilde{\chi}}(p_1)(a_LP_L+a_RP_R) e(p_2)\biggr] \biggl[\bar{e}(p_2^{\prime})(a_LP_L+a_RP_R)\tilde{\chi}(p_1^{\prime})(-ie)(p_1+p_2-r)^\mu \biggr] \biggl[{1\over 2(p_1+p_2)\cdot(-r)}\biggr] \nonumber \\
& & \biggl[{1\over (p_1+p_2)^2-m_{\tilde{e}}^2-im_{\tilde{e}}\Gamma(p_1+p_2)} -{1\over (p_1+p_2-r)^2-m_{\tilde{e}}^2-im_{\tilde{e}}\Gamma(p_1+p_2-r)}\biggr].
\eea
Therefore, $M_1 + M_2 + M_3 = 0$, as expected, and we complete the proof of the
independent gauge invariance for both the $s$- and the $u$-channel sets of
Feynman graphs.

\end{document}